\documentclass{emulateapj}
\usepackage{amsmath,amssymb,graphicx,hyperref,natbib}
\newcommand{\Mbin}{M_\mathrm{bin}}
\newcommand{\ah}{a_\mathrm{h}}
\journalinfo{The Astrophysical Journal, 810:49 (17pp), 2015 September 1}
\submitted{Received 2015 May 26; accepted 2015 July 27; published 2015 August 27}

\begin{document}

\title{The final-parsec problem in the collisionless limit}

\author{Eugene Vasiliev\altaffilmark{1,2}} \email{eugvas@lpi.ru}
\author{Fabio Antonini\altaffilmark{3}} \email{fabio.antonini@northwestern.edu}
\author{David Merritt\altaffilmark{4}} \email{merritt@astro.rit.edu}
\affil{$^1$Lebedev Physical Institute, Moscow, Russia}
\affil{$^2$Rudolf Peierls Centre for Theoretical Physics, Oxford, OX1 3NP, UK}
\affil{$^3$Center for Interdisciplinary Exploration and Research in Astrophysics (CIERA) and \\ 
Department of Physics and Astrophysics, Northwestern University,  Evanston, IL 60208, USA}
\affil{$^4$School of Physics and Astronomy and Center for Computational Relativity and Gravitation, \protect\\
Rochester Institute of Technology, Rochester, NY 14623, USA}

\begin{abstract}
A binary supermassive black hole loses energy via ejection of stars in a galactic nucleus, 
until emission of gravitational waves becomes strong enough to induce rapid  coalescence. 
Evolution via the gravitational slingshot requires that stars be continuously supplied to 
the binary, and it is known that in spherical galaxies the reservoir of such stars is quickly 
depleted, leading to  stalling of the binary at parsec-scale separations. 
Recent $N$-body simulations of galaxy mergers and isolated nonspherical galaxies suggest 
that this stalling may not occur in less idealized systems. 
However, it remains unclear to what degree these conclusions are affected by collisional 
relaxation, which is much stronger in the numerical simulations than in real galaxies.
In this study, we present a novel Monte Carlo  method that can efficiently deal with 
both collisional and collisionless dynamics, and with galaxy models having arbitrary shapes. 
We show that without relaxation, the final-parsec problem may be overcome only in triaxial 
galaxies. Axisymmetry is not enough, but even a moderate departure from axisymmetry is sufficient 
to keep the binary shrinking. We find that the binary hardening rate is always substantially 
lower than the maximum possible, ``full-loss-cone'' rate, and that it decreases with time, 
but that stellar-dynamical interactions are nevertheless able to drive the binary to coalescence 
on a timescale $\lesssim 1$~Gyr in any triaxial galaxy.
\end{abstract}

\section{Introduction}

Binary supermassive black holes (SBHs) are naturally formed in galaxy mergers, if both merging 
galaxies contain a central SBH. The heavy objects quickly sink to the center of 
the merger remnant due to dynamical friction and form a binary system. 
Subsequent evolution of the binary is driven by interaction with stars in the galactic nucleus, 
which are ejected by the slingshot mechanism \citep{Saslaw1974} if they arrive within a 
distance $\lesssim a$ from the binary center of mass, where $a$ is the semimajor axis of 
the binary orbit. As a result, the orbit shrinks (hardens), and if the binary becomes 
sufficiently hard, emission of gravitational waves (GWs) becomes the main source of energy loss, 
rapidly bringing the two SBHs to coalescence. 
The efficacy of this process, however, depends crucially on the supply of stars into the loss 
cone (the low-angular-momentum region of phase space in which stellar orbits can approach 
the binary), and if the reservoir is depleted, the binary nearly stalls, or at least its 
shrinking timescale may become much longer than the Hubble time \citep{BegelmanBR1980}. 

In idealized spherical galaxies, the only guaranteed mechanism of loss-cone repopulation is 
two-body relaxation. The problem of feeding stars into the loss cone of the binary has 
much in common with the similar problem for a single SBH, which was extensively studied in 
the 1970s in the context of spherical systems \citep[e.g.][]{FrankRees1976,LightmanShapiro1977}. 
These papers identified an important distinction between empty- and full-loss-cone regimes: 
in the former, the flux of stars is inversely proportional to the relaxation time and depends 
on the size of the loss cone only logarithmically, while in the latter the relaxation is so 
efficient that the supply of stars into the loss cone becomes independent of the relaxation time
and proportional to the size of the loss cone. 
Relaxation times in real galaxies are so long that they are nearly always in 
the empty-loss-cone regime. The conjectured stalling of the binary evolution has been labeled 
the ``final-parsec problem'' \citep{MilosMerritt2003a}.

On the other hand, in  non-spherical galactic potentials the angular momentum of stars changes 
not only by two-body relaxation, but also by large-scale torques \citep[][Chapter 4]{MerrittBook}. 
The orbital structure of non-spherical galaxies is rather diverse, and in triaxial galaxies 
there exist an entire class of centrophilic orbits, (e.g.\ box or pyramid orbits) that may attain 
arbitrarily low values of angular momentum without any  relaxation. 
These orbits were identified as a promising mechanism of loss-cone repopulation 
\citep[e.g.][]{NormanSilk1983,MerrittPoon2004,HolleySigurdsson2006}.
Similarly, in an axisymmetric potential only one component of angular momentum is conserved, 
and the number of stars that can enter the loss cone is much larger than in the spherical case 
\citep{MagorrianTremaine1999,Yu2002,VasilievMerritt2013}, although not as large as in a triaxial 
system.

Numerical simulations of the evolution of binary SBHs face an important difficulty: 
since the number of particles $N$ in a typical simulation ($\lesssim 10^6$) is several orders 
of magnitude smaller than the number of stars $N_\star$ in a real galaxy, it is necessary 
to properly understand the scaling laws. The rate of collisional evolution is inversely 
proportional to the relaxation time, which scales roughly as $N/\log N$, while collisionless 
effects are essentially independent of $N$.
Even if only the collisional effects play a role in the dynamics, the hardening rate scales 
differently in the empty- and full-loss-cone regimes. For small $N$ the system is in the latter 
regime and the hardening rate is nearly independent of $N$, while for large $N$ the hardening 
rate should drop with $N$. 
Early studies were restricted to rather low values of $N$ and hence did not find any 
$N$-dependence of the hardening rate \citep[e.g.][]{QuinlanHernquist1997,MilosMerritt2001}, 
while in more recent simulations of isolated spherical systems with larger $N$ the hardening 
rate was found to decline with $N$ \citep[e.g.][]{MakinoFunato2004,BerczikMS2005,MerrittMS2007}, 
although less steeply than the $N^{-1}$ dependence expected in the empty-loss-cone regime.
On the other hand, simulations that considered isolated triaxial \citep{BerczikMSB2006,
Berentzen2009} or axisymmetric \citep{KhanHBJ2013} systems, or started from a merger of 
two galaxies, which need not result in a spherical model, typically find no dependence of 
hardening rate on $N$.
This has been interpreted as a sign that the binary remains in the full-loss-cone regime due 
to efficient reshuffling of angular momenta of orbits by non-spherical torques.

In a previous paper \citep[hereafter Paper I]{VasilievAM2014}, 
\defcitealias{VasilievAM2014}{Paper I}
we reconsidered binary hardening in isolated galaxies with different geometries 
(spherical, axisymmetric and triaxial) using high-resolution $N$-body simulations with $N$ 
up to $10^6$. Somewhat surprisingly, we found that the hardening rates do depend on $N$ in 
all three cases, although they decline less rapidly with $N$ for non-spherical models. 
With the exception  of the highest-$N$ integrations, there was almost no difference between 
axisymmetric and triaxial models.
We also explored the possible contribution to the hardening rate from collisionless effects, 
by analyzing the properties of orbits in our models and estimating the draining rates of 
centrophilic orbits. We concluded that with  presently accessible values of $N$ it is 
difficult to disentangle collisional and collisionless effects in loss-cone repopulation.
In real galaxies, however, collisional effects are expected to play a much smaller role, 
so that it is hard to draw firm conclusions about the evolution of binary SBHs in 
non-spherical galaxies based on conventional $N$-body simulations.

In this study, we return to the topic and present a novel Monte Carlo method that can be used 
to model the evolution of galaxies hosting binary SBHs.
Our new algorithm contains an adjustable rate of relaxation, which can even be set to zero, 
yielding the collisionless limit.
We demonstrate that in this limit, triaxial systems have enough centrophilic orbits to 
maintain an adequate supply of stars into the loss cone, although the binary hardening rate is 
not as high as in the full-loss-cone regime and slowly declines with time. Nevertheless, 
for all reasonable values of the parameters, the coalescence time is shorter than the Hubble time.
We therefore conclude that the final-parsec problem is solved by triaxiality even in a purely 
collisionless stellar system. 
By contrast, in axisymmetric and spherical galaxies the hardening rate rapidly drops
in the absence of relaxation, meaning that in most galaxies the binary would never merge. 
Crucially, collisional relaxation in conventional $N$-body simulations overwhelms the depletion 
of the loss cone and completely changes the long-term behavior of the binary.

In section~\ref{sec:method} we describe the Monte Carlo method used in this study and compare 
it to previous similar approaches, while in section~\ref{sec:tests} we validate the method 
against a large suite of conventional $N$-body simulations with $N\le2\times10^6$. 
Section~\ref{sec:evol_isolated} is devoted to a detailed study of the evolution of isolated 
galaxy models, constructed initially as equilibrium configurations in spherical, axisymmetric 
and triaxial geometries. We present the results of Monte Carlo simulations and illustrate  
the trends found in long-term evolution with simple analytical arguments. 
Based on these results, we compute coalescence times for binary SBHs as a function of galaxy 
structure and initial parameters of the binary orbit. 
In section~\ref{sec:evol_merger} we conduct $N$-body simulations of mergers and compare them 
to Monte Carlo models.
Finally, in section~\ref{sec:discussion}, we summarize our important results
and compare them with previous work on the final-parsec problem.

As in \citetalias{VasilievAM2014}, we focus on purely stellar-dynamical processes.
For a broader picture, including the effects of gas-dynamical torques,
see Section 8.4 of \citet{MerrittBook} or the recent review of \citet{Colpi2014}.
Some preliminary results from the work presented here were described in \citet{Vasiliev2014b}.

\section{Method}  \label{sec:method}

\subsection{Definitions}

We consider the evolution of a binary SBH composed of two point masses, 
$m_1$ and $m_2$, which are on a Keplerian orbit with semimajor axis $a$ and eccentricity $e$.
The total mass of the binary, $\Mbin\equiv m_1+m_2$, is a small fraction ($10^{-2}-10^{-3}$) 
of the total mass of surrounding stellar distribution. The mass ratio of the binary is
$q\equiv m_2/m_1 \le 1$. 

A star passing at a distance $\lesssim a$ from the binary undergoes a complex scattering 
interaction and is ejected with a typical velocity $\sim \sqrt{G\Mbin/a}$ 
(the characteristic orbital velocity of the binary); 
for a hard binary this is higher than the average velocity of the stellar population, 
thus the star carries away energy and angular momentum from the binary. 
The precise definition of a hard binary varies among different studies; here we adopt that 
a binary is hard if its semimajor axis is smaller than 
\begin{align}  \label{eq:a_hard}
\ah \equiv \frac{q}{4(1+q)^2}\, r_\mathrm{infl} ,
\end{align}
where, in turn, the radius of influence $r_\mathrm{infl}$ is defined as the radius enclosing 
the mass in stars equal to twice the total mass of the binary. 
This definition depends on the evolutionary phase, as the slingshot process reduces the density
of stars in the galactic nucleus. The most rapid depletion occurs just after the binary formation, 
and after the binary becomes hard the depletion slows down considerably. 
For consistency with merger simulations, in which it is not possible to assign any particular 
value to $r_\mathrm{infl}$ before the two galactic nuclei have merged, we adopted the convention 
to evaluate the influence radius \textit{after} the hard binary has formed 
(see \citealt{MerrittSzell2006} for a discussion on different definitions of $r_\mathrm{infl}$). 

Another important quantity is the hardening rate $S\equiv d(1/a)/dt$. 
In what follows, we will frequently compare $S$ with the reference value $S_\mathrm{full}$ 
that would occur if the distribution of stars in phase space were not affected by the presence 
of the binary -- in other words, if the loss cone was ``full''. 
Unfortunately this value also does not have a commonly accepted definition.
For instance, if the stars were uniformly distributed in space, with density $\rho$ and 
with isotropically directed velocities all of the same magnitude $v$, then the hardening rate 
can be expressed as
\begin{align}  \label{eq:S_uniform}
S_\mathrm{uniform} = H_1 \frac{G\rho}{v} ,
\end{align}
where the dimensionless coefficient $H_1$ can be measured from scattering experiments 
\citep[e.g.][]{Quinlan1996,SesanaHM2006}, and has a value $H_1\approx 18$ in the 
hard-binary limit.
A more widely used definition applies to systems with a uniform density and a Maxwellian 
velocity distribution  with dispersion $\sigma$:
\begin{align}  \label{eq:S_maxwell}
f(x,v)  &\equiv 
  \rho\, \tilde f(v) = 
  \frac{\rho}{(2\pi\,\sigma^2)^{3/2}}\,\exp\left(-\frac{v^2}{2\sigma^2}\right) \,,\nonumber\\
S_\mathrm{Maxwell} &\equiv 
  \int_0^\infty dv\, 4\pi v^2\,\tilde f(v)\,S_\mathrm{uniform}(v) = 
  H\frac{G\rho}{\sigma},
\end{align}
where $H\equiv H_1\sqrt{2/\pi}\approx 14.5$. 

The assumption of a uniform-density background is clearly an oversimplification, and a more 
robustly defined quantity is obtained by averaging the single-velocity hardening rate 
(\ref{eq:S_uniform}) over the actual distribution function of stars $f(x, v)$ in the entire galaxy.
If we assume that the latter is isotropic (i.e., depends only on the energy $E$, but not 
on the angular momentum), then 
\begin{align}  \label{eq:S_isotr}
S_\mathrm{iso} &\equiv \int_0^\infty\! dv\, 4\pi v^2\,f(E)\,\frac{H_1\,G}{v} =
4\pi\,H_1\,G \int_{\Phi_0}^0 dE\,f(E) \,.
\end{align} 

Here $\Phi_0$ is the depth of potential well of the stellar cusp (excluding the potential 
of the SBH), so that the integration includes energies corresponding 
to stars that are unbound to the binary, but still bound to the entire galaxy
(this is a rather ad hoc convention, but it is justified by the fact that few 
stars remain bound to the binary for any significant time after it becomes hard).
Another derivation of this quantity from a somewhat different perspective can be found e.g.\ in 
\citetalias{VasilievAM2014}, Equations~(7)--(9), or in Section~8.3 of \citet{MerrittBook}.
Note that the above value has no relation to the total mass of stars in the galaxy, which is 
given by $\int dE\,f(E)\,g(E)$, where $g(E)$ is the density of states and rapidly rises with $E$.

Yet another way to define a reference hardening rate is to substitute
the density and velocity dispersion evaluated at $r_\mathrm{infl}$ into Equation~(\ref{eq:S_maxwell})
\citep{SesanaKhan2015}. This definition has the advantage of being easily computed from
quantities that are  accessible to an $N$-body snapshot, without the need to calculate the 
distribution function; on the other hand, it reflects only the local parameters near the galaxy center
and not the entire population of stars that can interact with the binary.
Since the velocity dispersion can itself be expressed as 
$\sigma^2 \sim G\Mbin/r_\mathrm{infl}$, we may write
\begin{align}  \label{eq:S_infl}
S_\mathrm{infl} \equiv \mathcal A\,(G\Mbin)^{1/2}\,r_\mathrm{infl}^{-5/2} ,
\end{align}
which agrees quantitatively with the if 
Here the dimensionless factor $\mathcal A$ can be chosen to match the previous definition 
(\ref{eq:S_isotr}); its value depends weakly on galaxy structure and lies in the range 3--5.

In what follows, we will refer to either $S_\mathrm{iso}$ or $S_\mathrm{infl}$ as 
the ``full-loss-cone hardening rate''. We stress that there is no fundamental reason to expect that 
the actual hardening rate will be close to $S_\mathrm{full}$, since the distribution of 
stars at low angular momenta is not isotropic, and even the value of $S_\mathrm{full}$ 
need not remain constant in time -- it merely serves as a useful reference value.

The loss of energy and the evolution of eccentricity due to the emission of GWs 
are given by the following expressions \citep{Peters1964}:
\begin{align}  \label{eq:hardening_GW}
\frac{d(1/a)}{dt} &=  \frac{1/a}{T_\mathrm{GW}}\,,
  \quad T_\mathrm{GW}\equiv \frac{5}{64} \frac{c^5 a^4}{G^3M^3} \frac{(1+q)^2}{q} f(e) , \\
f(e) &\equiv \frac{(1-e^2)^{7/2}}{1+\frac{73}{24}e^2+\frac{37}{96}e^4} , \label{eq:f_ecc} \\
\frac{d e^2}{dt} &= -\frac{G^3 M^3}{c^5 a^4}\frac{q}{(1+q)^2} 
  \frac{2e^2 (304+121e^2)}{15(1-e^2)^{5/2}} . \label{eq:ecc_GW}
\end{align}

Throughout the paper, we will mostly present the results in dimensionless $N$-body units, 
with the mass and the scale radius of the galaxy model both set to unity. The models may be 
scaled to a given galaxy using any two out of three fundamental scales (length, time and mass). 
To simplify the discussion, we reduce this freedom to a one-parameter family of galaxies 
in which the length and mass scales are related through the $M_\bullet-\sigma$ relation 
in the following form \citep[e.g.][Fig.12]{MerrittSK2009}:
\begin{align}  \label{eq:Msigma}
r_\mathrm{infl} = r_0 \times (M_\bullet/10^8\,M_\odot)^\kappa, \quad r_0=30\mbox{ pc},\; \kappa=0.56.
\end{align}

\subsection{Monte Carlo code}

The Monte Carlo method used in this work is an extension of the \textsc{Raga} code 
\citep{Vasiliev2015}. 
We follow the motion of $N$ particles in the combined potential of the stellar distribution,
$\Phi_\star(\boldsymbol{r})$, and the two point masses orbiting each other, centered at origin.
The orbit of the binary is assumed to be Keplerian, aligned in the $x-y$ plane, and elongated 
in $x$ direction; we make no attempt to follow either the change of the orbital plane, which 
we know to be small from the $N$-body simulations (although it could be quite significant
over long timescale in triaxial galaxies, see \citealt{CuiYu2014}), or the precession of its 
periapsis, which we assume is not particularly important for dynamics. 
The evolution is broken down into many small intervals of time (episodes) of duration 
$T_\mathrm{epi} \gg T_\mathrm{bin}$; during each episode we keep the stellar potential and 
the parameters of the binary orbit unchanged. 

Each particle is moving independently from the others under the influence of three forces: 
the gradient of the smooth static stellar potential, represented as a spherical-harmonic 
expansion with spline-interpolated coefficients as functions of radius; 
the time-dependent force from the binary; and random velocity kicks that model the effect 
of two-body encounters for a system composed of a certain number of stars, $N_\star$, which 
does not need to be related to the number of particles in the simulation, and can even be 
set to infinity (thus switching off the two-body relaxation).
The velocity perturbations are computed from the local (position-dependent) drift and diffusion 
coefficients, using the standard formalism from the relaxation theory \citep[e.g.][Chapter 5]
{MerrittBook} and an isotropic approximation for the distribution function of stars (i.e.\ 
with dependence on energy only).
Thus our method descends from the Spitzer's formulation of Monte Carlo approach 
\citep{SpitzerHart1971} and does not rely on orbit-averaging, as other contemporary Monte Carlo 
codes that use the formalism of \citet{Henon1971}. 

During the encounter of a particle with the binary, defined as the time when the distance 
of the particle from origin is less than $10a$, we record the changes in energy and angular 
momentum of the particle that arise due to the motion in time-dependent potential of the binary
(not including the perturbations from relaxation or the torque from the non-spherical stellar 
potential). 
After each particle's trajectory has been calculated over the entire episode,
we sum up these changes for all particles that experienced one or more encounters with 
the binary, and change the binary's energy and angular momentum by the same amount with 
the opposite sign, while keeping its orbital phase unchanged from the end of the episode 
to the start of the next one. Thus the stars do not exert force on the black holes directly, 
but we use conservation laws to impose the changes to the binary orbit parameters.
The binary orbit parameters are optionally modified after each episode according to 
the expressions for gravitational-wave emission (\ref{eq:hardening_GW},\ref{eq:ecc_GW}).
The stellar potential and the diffusion coefficients that account for two-body relaxation 
are also updated at the end of each episode, reflecting the changes in the stellar distribution 
in the course of evolution (here the most important effect is the gradual erosion of the central 
stellar cusp).

\subsection{Comparison with similar approaches}

It is instructive to compare our method with the other schemes used for an approximate treatment 
of the joint evolution of the binary and the stellar distribution.
\citet{QuinlanHernquist1997} developed a program \textsc{scfbdy} which combines elements from 
the self-consistent field (SCF) method \citep{HernquistOstriker1992} with a direct integration 
of black hole--star interactions. The two black holes were integrated using \textsc{nbody2}
\citep{Aarseth1999} -- a direct-summation code with neighbor scheme, adaptive timestep, and 
optional two-body regularization, taking into account the forces from all stars individually. 
The motion of star particles, on the other hand, was computed in the gravitational potential 
of the two black holes and the smooth potential of the other stars, represented by a basis-set 
expansion \citep{HernquistOstriker1992}. 
Their approach is quite similar to ours, with the following differences:
(a) the binary's center of mass is not fixed at origin, although the center of stellar 
potential expansion is;
(b) the star particles exert ``real'' forces on the black holes, which lead directly to the 
changes in the binary orbit, instead of relying on the Newton's third law as in our method;
(c) two-body relaxation is not modeled.
They used a rather small update interval for the stellar potential, which could lead to 
artificial relaxation due to fluctuations in the expansion coefficients 
\citep{HernquistBarnes1990,Weinberg1996,Sellwood2015}; in our code we recompute the potential 
less frequently and use $N_\mathrm{samp}\gg 1$ sampling points from each particle's trajectory 
stored during each episode, to further reduce discreteness noise. 
Nevertheless, any finite-$N$ system is not entirely free of relaxation \citep[e.g.][]
{Weinberg1998}, so our simulations without explicitly added two-body relaxation should be 
regarded as an upper limit for the evolution rate expected in truly collisionless systems.
Finally, \citet{QuinlanHernquist1997} only dealt with nearly spherical systems, 
although in principle their scheme can equally well work for non-spherical systems after 
switching on the corresponding terms in the spherical-harmonic expansion (as in our method).

\citet{HemsendorfSS2002} used a similar approach in their code \textsc{EuroStar}: the two 
black holes and a subset of star particles with angular momenta lower than a certain threshold 
are integrated with a direct-summation method, adapted from Aarseth's \textsc{nbody6} code, 
while the rest of the star particles are followed in the smooth field by a modified version 
of the SCF method. 

\citet{SesanaHM2006,SesanaHM2008} used a large suite of three-body scattering 
experiments to derive the expressions for the rate of hardening and eccentricity growth, 
extending the results of \citet{Quinlan1996} to a wider range of mass ratios and eccentricities. 
\citet{SesanaHM2007} and \citet{Sesana2010} developed a hybrid approach for the evolution of 
the binary, combining the analytical fits to these scattering experiments with a time-dependent 
model for the loss cone draining and repopulation. 
They did not simulate the effect of two-body relaxation and Brownian motion of the binary 
center of mass explicitly, but included some prescriptions to take it into account. 
The changes in the stellar potential due to ejection of stars were ignored.

\citet{MeironLaor2012} introduced another scheme based on the conservation laws to find 
the changes in the binary's orbit.
The motion of stars in the static spherically symmetric potential of the stellar cusp 
plus the time-dependent potential of the binary is followed over a short interval of time,
after which the changes in the total energy and angular momentum of all stars are translated 
into the forces that should be applied to the black holes.
Unlike our method, in which we assume a Keplerian orbit for the binary and adjust its parameters 
on a timescale much longer than the orbital period, their approach requires rather short 
update intervals to follow the motion of the black holes directly. 

Compared to the previous studies, our approach most closely resembles that of 
\citet{QuinlanHernquist1997} and \citet{HemsendorfSS2002}, in that we also use 
the spherical harmonic expansion technique to represent the smooth potential of the stellar 
cluster, which is itself a nearly collisionless system, while considering the interaction 
between stars and the binary as a succession of three-body scattering events, as in 
the series of papers by Sesana et al. 
The treatment of the binary is somewhat more approximate in our method -- we assume its 
center-of-mass to be at rest, and ignore the changes in its orbital plane and orientation 
of the line of nodes. The evolution of its most important parameters -- binding energy and 
eccentricity -- is deduced from the orbits of stars using conservation laws, in a similar 
manner to \citet{MeironLaor2012}, but using much longer update intervals, spanning many 
orbital periods. The most important new feature of our method with respect to the previous 
studies is the ability to efficiently model the two-body relaxation with adjustable magnitude, 
allowing us both to compare our Monte Carlo simulations with direct $N$-body simulations 
(using the same relaxation rate), and to extend them into nearly collisionless regime, 
by switching off the relaxation. 

Fokker--Planck methods have also been applied to this problem, always in the spherical
geometry \citep{MilosMerritt2003b, MerrittWang2005, MerrittMS2007}.
The Fokker--Planck approach has the advantage of a potentially much finer resolution of 
the loss-cone region in $(E,L)$ space. A disadvantage is the need to orbit-average 
the diffusion coefficients. Our Monte Carlo algorithm avoids that limitation; on the other 
hand, the orbit-averaged approximation only breaks down near the loss cone, and at least 
in the spherical geometry, a boundary-layer treatment of the loss cone is available that is
based on the local (not orbit-averaged) Fokker--Planck equation and which automatically 
accounts for empty vs.\ full loss cones \citep{CohnKulsrud1978}.
The secondary slingshot is also more naturally accounted for in a Monte Carlo algorithm,
although it has also been implemented in Fokker--Planck treatments \citep{MerrittMS2007}.
The primary advantage of our method over Fokker--Planck algorithms is the ability to model 
stellar systems regardless of their shape, including the chaotic orbits that arise in 
non-spherical geometries.
Limitations of our approach include neglect of Brownian motion (justified to some extent below 
in Section~\ref{sec:evol_longterm_theory}) and the assumption of near-equilibrium of 
the stellar distribution, which means that it is not expected to perform well in highly dynamic 
situations such as mergers.
It is also worth noting that the method scales linearly with $N$ and already at $N=10^5$ 
outperforms the GPU-accelerated direct-summation code by a factor of 20.

\section{Tests}  \label{sec:tests}

We first checked that the results of our calculations do not depend on the technical 
parameters such as the number of radial grid points in the spherical-harmonic expansion, 
the update interval, or the type of orbit integrator and its accuracy parameters, 
provided that they are set to reasonable values. 

We also verified that the conservation-law method produces the same results as the scattering 
experiments of \citet{Quinlan1996} and \citet{SesanaHM2006} which used a small but non-zero 
mass of the incoming star, and computed its effect on the binary orbit directly. 
In particular, the hardening rates and eccentricity growth rates were found to be comparable 
to Figures 3 and 4 of \citet{SesanaHM2006} for a uniform-density isothermal background 
population of stars.

Next we compared the Monte Carlo code with a large suite of conventional $N$-body integrations 
of isolated galaxy models with various shapes and initial parameters for the binary. 
These were similar to the simulations we used in \citetalias{VasilievAM2014}, but covered 
a wider range of eccentricities, mass ratios, and density profiles. 
The initial density profile of stars follows the \citet{Dehnen1993} model with the inner cusp 
slope $\gamma=1$ (our default model) and $\gamma=2$, in three different geometries:
spherical, oblate axisymmetric (axis ratio $1:0.8$) and triaxial (axis ratios $1:0.9:0.8$ -- 
our default model, -- and $1:0.8:0.6$). 
These are constructed to be in equilibrium with a central SBH of mass $M_\bullet=0.01$ 
of the total stellar mass, and have a nearly isotropic velocity distribution. 
The mass $M_\bullet$ is split between two components of the binary, and the two SBHs 
are initially placed at a separation 0.2 (for $\gamma=1$) or 0.02 (for $\gamma=2$), 
slightly larger than the radius of influence.%
\footnote{Recall that we measure $r_\mathrm{infl}$ just after the hard binary has formed; 
for instance, in a $\gamma=1$ Dehnen model with a single SBH $r_\mathrm{infl}=0.165$, 
but after the binary becomes hard, the density cups is eroded and $r_\mathrm{infl}$ increases 
to $\sim0.2$ for the models with $q=1$, changing only slightly in the subsequent evolution.}
We considered two values for the mass ratio -- $q=1$ (equal-mass binary) and $q=1/9$.
Using (\ref{eq:Msigma}), our $\gamma=1,q=1$ models can be scaled to real galaxies so that one 
length unit equals $150\;\mbox{pc}\times(M_\bullet/10^8\,M_\odot)^{0.56}$ and one time unit 
is $0.27\;\mbox{Myr}\times(M_\bullet/10^8\,M_\odot)^{0.84}$.

The $N$-body simulations were performed with the direct-summation code $\phi$GRAPEch 
\citep{HarfstGMM2008}, using the \textsc{sapporo} library for GPU acceleration 
\citep{GaburovHP2009}. 
This code employs chain regularization to accurately follow the motion of the binary 
and the stars interacting with it, in exact Newtonian gravity (no softening); 
we used a very small softening $\epsilon=10^{-6}$ for particles outside the chain.
Thanks to the use of the chain, the relative error in energy was typically below $10^{-5}$ 
at the later stages of evolution, when both massive particles are included in the chain.

The initial conditions for Monte Carlo models were taken from $N$-body simulations 
at the time when the binary has just become hard, reaching the semimajor axis $\ah$.
The relaxation rate is a free parameter in the Monte Carlo method, defined by the number 
of stars in the target system $N_\star$ and the Coulomb logarithm $\ln\Lambda$.
We have measured the changes in energies and angular momenta of particles in the spherical 
$N$-body simulations, and compared them to the expected diffusion rate as a function of energy
\citep[see][figure~1]{Vasiliev2015}; the value of $\Lambda=0.02N_\star$ provided the best 
agreement for $10^5\lesssim N_\star\lesssim 10^6$. The prefactor in the Coulomb logarithm 
is the only free parameter in the code that could be assigned at will (apart from technical 
parameters such as the number of terms in the potential expansion), and we have adopted the above 
definition to match the diffusion rate, not the hardening rate of the binary or anything else; 
thus if the other aspects of evolution agree between Monte Carlo and $N$-body simulations, 
this demonstrates the predictive power of the Monte Carlo approach.

\begin{figure}
\includegraphics{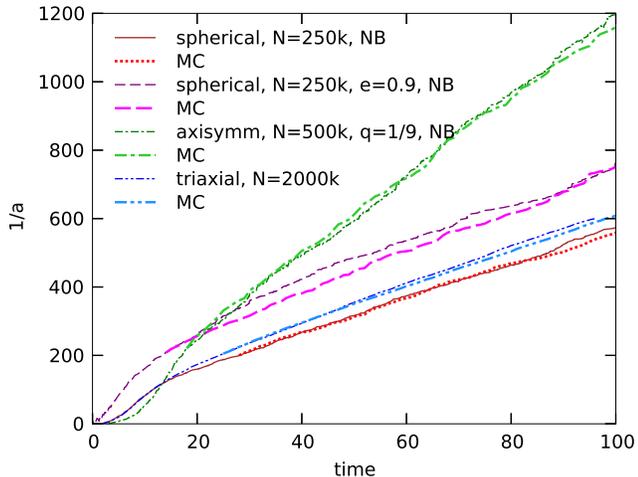}
\caption{
Evolution of inverse semimajor axis as a function of time, for a few models with various $N$, 
initial eccentricity, binary mass ratio $q$, and geometry.
Thinner lines are from $N$-body simulations, and thicker ones are from Monte Carlo simulations.
} \label{fig:mc_nb}
\end{figure}

\begin{figure}
\includegraphics{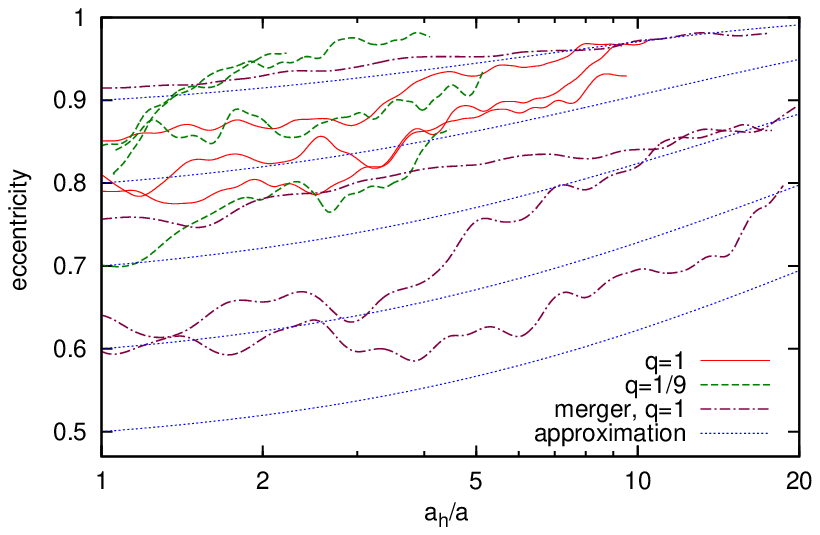}
\includegraphics{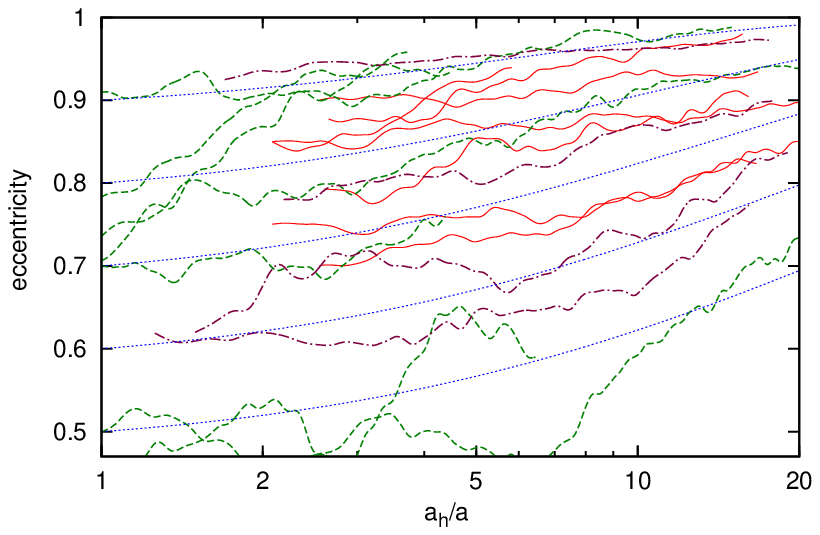}
\caption{
Evolution of binary eccentricity as a function of separation, in $N$-body (top panel) and 
Monte Carlo (bottom panel) simulations.
Shown are tracks for isolated systems with equal-mass binaries (solid red) and 
$q=1/9$ (dashed green), and merger simulations (dot-dashed purple);
the adopted analytic prescription for the evolution of eccentricity (Equation~\ref{eq:K_ecc}) 
is shown by the blue lines.
} \label{fig:K_ecc}
\end{figure}

We compared the Monte Carlo and $N$-body simulations using a number of criteria. 
The most important are the binary parameters -- semimajor axis and eccentricity.
It should be noted that individual simulations have a considerable scatter in the hardening 
rates and the evolution of eccentricity, especially at low $N$. Nevertheless, statistically 
the agreement between Monte Carlo and $N$-body simulations is very good for a wide range of 
parameters ($N$, initial eccentricity, geometry, binary mass ratio), see 
Figure~\ref{fig:mc_nb} for a few examples. 
We have also checked that the long-term behavior of Monte Carlo simulations with the same 
relaxation rate (set by $N_\star$) but with different number of particles $N$ is similar; 
this allows to extrapolate our method to large $N_\star$ while still using a reasonable 
($N\lesssim10^6$) number of particles.

The dependence of hardening rate on eccentricity was found to be weak; if anything, models 
with high $e$ evolved a little faster, in agreement with the results of scattering experiments 
of \citet{SesanaHM2006}, \citet{MerrittMS2007}, but typically the difference was comparable 
to the scatter between individual runs. 
The evolution of eccentricity itself is in fact more important, since the energy losses due to 
GW emission depend strongly on $e$. 
Scattering experiments typically suggest a slow but steady growth of eccentricity, 
but in practice its evolution is more erratic, because individual interactions may both 
increase and reduce $e$ (thus the outcome results from a slight imbalance between them).
Stars with smaller angular momenta tend to reduce $e$ \citep{SesanaHM2008}, and so do stars 
that corotate with the binary \citep{Iwasawa2011, SesanaGD2011}. In our simulations, the fraction 
of corotating and counterrotating stars is approximately equal, so the latter factor does not 
come into play, but the distribution of stars in angular momentum does depend on the details 
of loss-cone repopulation, so the former effect is quite important.
The eccentricity itself stayed roughly constant if it was initially small, and growed slightly 
if started from $e\gtrsim 0.5$. 
Models with initial $e\ge 0.8$ typically increased it to $0.9\lesssim e \lesssim 0.95$ 
in the course of evolution, especially in the unequal-mass case ($q=1/9$). 
Nevertheless, neither in $N$-body nor in Monte Carlo simulations did we observe a rapid 
growth of eccentricity to values $\gtrsim 0.99$, found in some previous studies 
\citep[e.g.][]{Iwasawa2011, MeironLaor2012}, although we did not consider systems with such 
high mass ratio as in the former study.

The eccentricity growth is traditionally described with a dimensionless parameter 
$K \equiv de/d\ln(1/a)$. Previous studies have found that in the hard-binary limit, 
$K$ reaches a maximum value of $\sim 0.1-0.2$ at $e\simeq 0.7$, and drops to zero at $e=0$ or 1.
We adopt the following parametrization for $K$, which is comparable to the findings of 
\citet{Quinlan1996} and \citet{SesanaHM2006}:
\begin{align}  \label{eq:K_ecc}
K &\equiv \frac{de}{d\ln(1/a)} \approx A\,e\,(1-e^2)^{b} \left(1+\frac{a}{a_0}\right)^{-1} ,\\
  &b=0.6,\; a_0=0.2 \ah,\; A=0.3 \,. \nonumber
\end{align}

Figure~\ref{fig:K_ecc} compares the evolution of eccentricity based on the above equation 
(dotted lines) to the results of $N$-body and Monte Carlo simulations. 
Even though there is considerable scatter between runs, the overall trend is reasonably well 
described by the fitting formula (\ref{eq:K_ecc}).

\section{Evolution of isolated galaxy models}  \label{sec:evol_isolated}

\subsection{Long-term evolution of stellar-dynamical hardening rate}  \label{sec:evol_longterm_mc}

\begin{figure}
\includegraphics{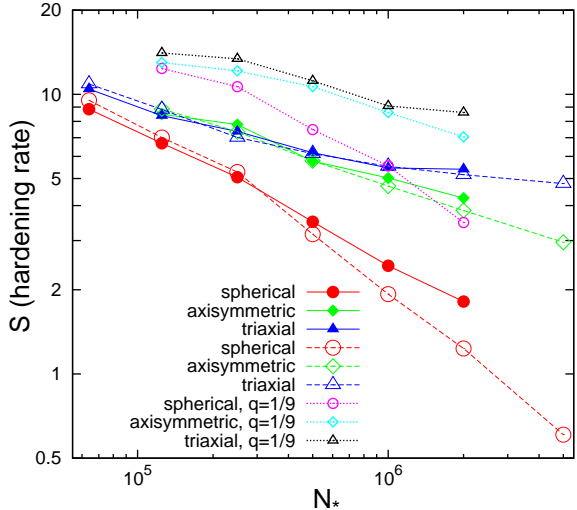}
\caption{
Hardening rates as functions of $N_\star$ for different geometries and mass ratios, 
computed on the interval $50\le t\le 100$ from $N$-body simulations 
(filled symbols, solid lines) and Monte Carlo simulations (open symbols, dashed lines).
The agreement is quite good, but Monte Carlo models underestimate the hardening rate
for $N_\star\ge 10^6$ in the spherical case, presumably due to the neglect of Brownian motion.
} \label{fig:hardening_rate}
\end{figure}

\begin{figure}
\includegraphics{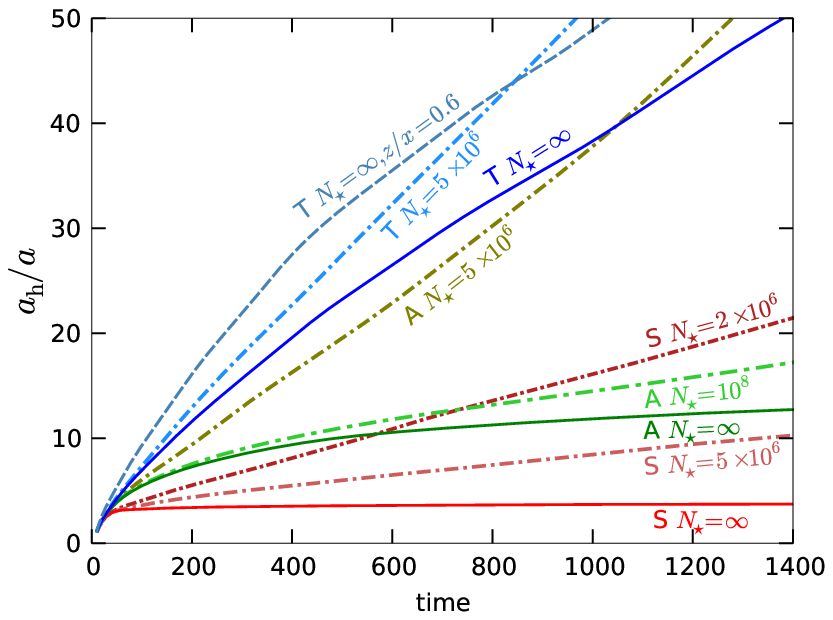}
\includegraphics{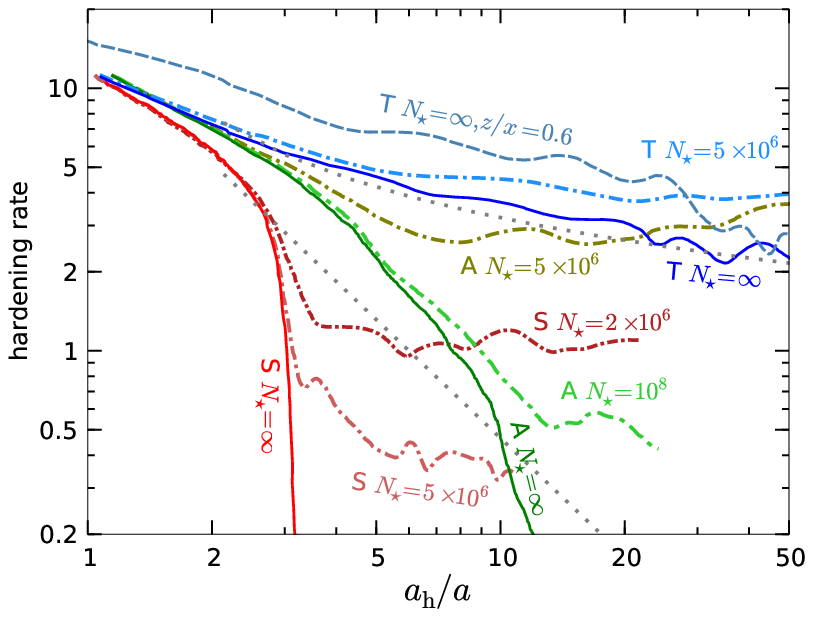}
\caption{
Long-term evolution of binary hardness for the equal-mass case and three different geometries
(\textbf{S}pherical, \textbf{A}xisymmetric and \textbf{T}riaxial).
Shown are curves corresponding to different relaxation rates (dash-dotted lines: 
$N_\star=2\times10^6, 5\times10^6$ and $N_\star=10^8$) and to the simulations with no relaxation 
(solid lines), dashed line is for the triaxial model with no relaxation and stronger initial 
flattening ($y/x=0.8,z/x=0.6$, while the other models have $z/x=0.8$ and, in the triaxial case,
$y/x=0.9$).
It is apparent that in the spherical and axisymmetric cases the binary separation approaches 
an upper limit without relaxation, while in the triaxial case it continues to shrink. Also notable 
is that even a modest amount of relaxation keeps the binary from stalling, although the evolution 
rate could be quite low for a realistic number of stars (even for a very low-mass binary, 
$\Mbin=10^6\,M_\odot$, the model scaled to a real galaxy would have $N_\star=10^8$). \protect\\
Top panel shows the time evolution of semimajor axis, and bottom panel shows the dependence 
of hardening rate on $\ah/a$. The full-loss-cone rate (\ref{eq:S_isotr}) is around 20 in our models,
higher than any value measured in the simulations.
Models with relaxation eventually attain a nearly constant hardening rate, depending on 
$N_\star$ and geometry. In collisionless simulations, by contrast, the hardening rates keep 
decreasing, very mildly in the triaxial case and steeply in the axisymmetric case. 
Dotted lines show the asymptotic expressions for the hardening rate in scale-free galaxies 
(\ref{eq:hardening_rate_asympt_triax},\ref{eq:hardening_rate_asympt_axisym}). 
} \label{fig:longterm}
\end{figure}

We now consider the long-term evolution of binary SBH in galaxies with a realistically large 
number of stars $N_\star$. 
Figure~\ref{fig:hardening_rate} shows the hardening rates $S$ computed on the interval $50\le t 
\le100$ for a series of $\gamma=1$ models with different $N_\star$, shape and binary mass ratio. 
It appears that in both spherical and axisymmetric cases the hardening rates continue to drop 
with increasing $N_\star$, but triaxial models tend to a nonzero limiting value of $S$ as 
$N_\star\to\infty$. 
This was already suggested in \citetalias{VasilievAM2014} based on $N$-body simulations, but 
the number of particles $N\le 10^6$ was not enough to establish it clearly; additional 
simulations with $N=2\times10^6$ and Monte Carlo models support this conclusion. 

Figure~\ref{fig:longterm} shows simulations extended to a much longer interval of time, 
approximately 0.4~Gyr for a model scaled to a galaxy with $\Mbin=10^8\,M_\odot$.
For each geometry we show the simulation without relaxation (i.e.\ in the collisionless limit), as 
well as for a moderately small relaxation rate (smaller than is achievable in $N$-body simulations).
It is immediately clear that in the collisionless limit there is a striking difference between 
three geometries: in the spherical case, the binary stalls at a semimajor axis barely smaller than 
$\ah$, and in the axisymmetric case it shrinks roughly a factor of ten below $\ah$, but ultimately 
also stalls. By contrast, in the triaxial case the binary continues to shrink, although 
the hardening rate decreases with time. 
In the simulations that include relaxation, however, the situation is quite different -- 
the binary never stalls in any geometry. For the triaxial case, the hardening rate in the 
$N_\star=5\times10^6$ simulation is already not much higher than in the collisionless limit,
but the axisymmetric systems differ dramatically from their collisionless counterpart: in both 
$N_\star=5\times10^6$ and $N_\star=10^8$ systems the hardening rate decreases at early times, 
as in the collisionless limit, but unlike the latter, it then attains a non-zero lower limit.

\begin{figure}
\includegraphics{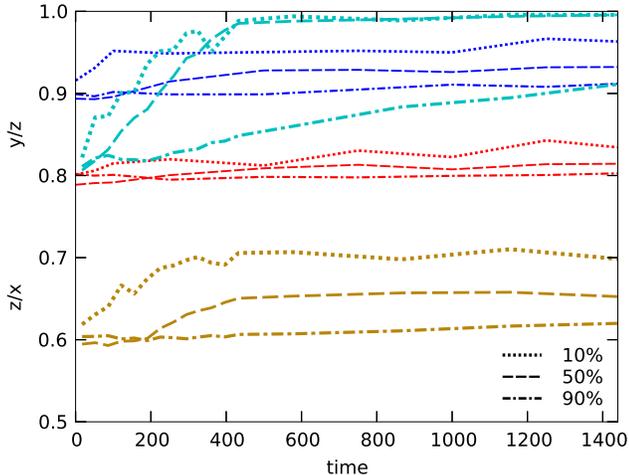}
\caption{
Evolution of shape of collisionless triaxial models. 
Dotted, dashed and dash-dotted lines show the axis ratio ($y/x$ -- top curves, blue and cyan,
$z/x$ -- bottom curves, red and gold), measured at radii enclosing 10, 50 and 90\% 
of total mass for the models with initial $x\!:\!y\!:\!z=1\!:\!0.9\!:\!0.8$ 
(thinner lines) and $1\!:\!0.8\!:\!0.6$ (thicker lines with longer dashes). 
The latter model evolves much quicker toward axisymmetry in the central parts.
} \label{fig:shape}
\end{figure}

We also considered a model with a stronger initial degree of triaxiality (axis ratio 
$1\!:\!0.8\!:\!0.6$); 
initially it had a higher hardening rate, which then dropped to a level comparable to that of 
the less flattened model. While a higher hardening rate is naturally explained by a larger 
reservoir of chaotic orbits in a more triaxial model (see next section), its subsequent decrease 
is a result of the loss of triaxiality in the central parts of the model (Figure~\ref{fig:shape}).
Interestingly, the model with a milder initial flattening retained its shape better.
This evolution toward axisymmetry clearly occurs due to the binary: if we replace it with 
a single SBH, the shape stays nearly constant on much longer intervals of time
than considered here.
It is commonly assumed that even single SBHs inevitably destroy triaxiality 
\citep[e.g.][]{GerhardBinney1985}.
\citet{MerrittQuinlan1998} simulated growth of black holes in triaxial models formed via cold 
collapse and found that central SBHs containing more than $\sim 3\%$ the mass of their host 
galaxies induced evolution toward axisymmetry.
However, later studies have shown that this does not necessary happen in isolated stationary systems:
using a more flexible, orbit-superposition algorithm, \citet{PoonMerritt2002} found
that self-consistent and apparently stable models could be constructed in which the SBH
mass was a substantial fraction of the nuclear mass, a result confirmed in other studies
\citep{HolleyBockelmann2002,PoonMerritt2004}.
\citet{Vasiliev2015} demonstrated, using the same Monte Carlo code, that relaxation is the main 
driving force behind shape evolution: in the collisionless regime the shape remains nearly constant. 
The likely reason is that the diffusion of chaotic orbits, which tends to erase triaxiality, 
is greatly facilitated by the noise from two-body relaxation \citep[e.g.][]{KandrupPS2000}. 
The shape evolution seen in the collisionless simulations with a binary SBH  -- as opposed to 
a single SBH -- may be caused by resonant perturbations of chaotic orbits by the time-dependent 
gravitational field \citep{KandrupSTB2003}, although more work is needed to explore this effect.

In addition, we performed simulations of models with a steep cusp ($\gamma=2$) and otherwise 
the same parameters. They demonstrated a similar behavior, with only the triaxial model 
continuing to shrink indefinitely in the collisionless case, although the hardening rate was 
dropping with time more rapidly than in the $\gamma=1$ model.
The loss of triaxiality was mild, much like the $\gamma=1$ model with the same axis ratios 
($1\!:\!0.9\!:\!0.8$).

Before we proceed to a theoretical explanation of the hardening rates, we discuss the properties 
of orbits in our models. 

\subsection{Properties of orbits interacting with the binary}  \label{sec:orbits}

We have examined the types of orbits that bring stars into interaction with the binary.
For each particle that arrives at a distance less than $10a$ from the center of mass of 
the binary, we record its phase-space coordinates at its first approach.
If a particle is scattered more than once by the binary, we only consider the first interaction, 
because it is difficult to define unambiguously when one interaction ends and the next one begins.
Then we compute a trajectory starting from these initial conditions in a smooth stellar potential 
plus a \textit{single} point mass $\Mbin$ at the origin, for a time corresponding to 100 orbital periods. 
This allows us to use powerful analysis tools applicable to orbits in smooth stationary potentials.
To distinguish regular and chaotic orbits, we use the Lyapunov exponent, and to determine 
if an orbit is centrophilic, we follow the algorithm described in the appendix of 
\citetalias{VasilievAM2014}.

After the binary became sufficiently hard ($a\lesssim 0.3\,\ah$), most of the orbits that 
interact with it are unbound to the binary (have energies higher than the depth of stellar 
potential well $\Phi_0$).
In the non-spherical cases, almost all such orbits ($\gtrsim 90\%$) are found to be chaotic, 
and for the triaxial system, a similarly large fraction of them are truly centrophilic, 
i.e.\ may attain arbitrarily low values of angular momentum. There are no genuinely centrophilic 
orbits in axisymmetric systems, because the variation of angular momentum is bounded from below 
by its conserved $z$-component, but for chaotic orbits the average value of $L$ is typically 
much larger than its minimum achievable value $L_z$, thus they constitute a reservoir of 
``usable'' orbits (loss region) that is much larger than the loss cone itself.

Regarding the \textit{overall} orbital structure of the model, the chaotic orbits are a minority 
($\sim 10\%$, depending on the degree of flattening and triaxiality). 
As discussed above, the shape of triaxial models gradually evolves toward axisymmetry from 
the inside out, but globally it remains sufficiently triaxial to support a substantial population 
of centrophilic orbits -- their fraction is roughly proportional to the deviation from axisymmetry, 
and their total mass is thus considerably larger than the mass of the binary.

\subsection{Theoretical models for the hardening rate evolution}  \label{sec:evol_longterm_theory}

The evolution of binary in the collisionless limit can be qualitatively described with a rather 
simple model for the draining of the population of orbits that can interact with the binary -- 
similar to the one presented in \citetalias{VasilievAM2014}, but without a detailed analysis of 
properties of orbits in a particular simulation.

We begin with the triaxial case, and assume that at each energy there is initially a fixed fraction 
$\eta$ of chaotic centrophilic orbits, which occupy the low-angular-momentum region of phase space
$L^2\le \eta L_\mathrm{circ}^2(E)$.
To account for their gradual depletion, we introduce the fraction of surviving orbits $\xi(E,t)$, 
so that the mass of chaotic orbits is 
\begin{align}  \label{eq:Mch_triax}
M_\mathrm{ch} = \int_{\Phi_0}^0 
  dE\,4\pi^2\,T_\mathrm{rad}(E)\,\eta L_\mathrm{circ}^2(E)\,\xi(E,t)\,f(E)\,,
\end{align}
where we have neglected the $L$-dependence of radial orbital period $T_\mathrm{rad}$.
The hardening rate due to interaction with these low-angular-momentum orbits is given by 
a generalization of Equation~\ref{eq:S_isotr}:
\begin{align}  \label{eq:hardening_rate_xi}
S(t) \equiv \frac{d(1/a)}{dt} = 4\pi\,H_1\,G \int_{\Phi_0}^0 dE\,f(E)\,\xi(E,t) .
\end{align}

On the other hand, the same interactions eject stars and decrease the fraction of surviving orbits.
Equating the energy carried away by stars with the change in the binary's binding energy gives
\begin{align}
\frac32 \frac{G\mu}{a} dM_\mathrm{ch} \approx -\frac{Gm_1m_2}{2} d\left(\frac{1}{a}\right)
\end{align}
with $\mu\equiv m_1m_2/M_\mathrm{bin}$ is the binary reduced mass 
\citep[e.g.][equation~25]{Merritt2004}. Thus
\begin{align}
\frac{dM_\mathrm{ch}}{dt} \approx 
-\frac13 \frac{m_1m_2}{\mu}\, a\, \frac{d}{dt}\left(\frac{1}{a}\right) \approx
-\frac13 M_\mathrm{bin\,} a\,S .
\end{align}

If we assume that the stars interacting with the binary become unbound to the galaxy (i.e., neglect 
the secondary slingshot), then the draining of the loss region occurs independently at each  
energy, and we can write  the following equation for the evolution of surviving fraction of stars:
\begin{align}  \label{eq:depletion_triax}
\frac{d\xi(E,t)}{dt} = -\xi(E,t) \frac{H_1 }{6\pi} 
  \frac{2G\Mbin a(t)}{\eta L_\mathrm{circ}^2(E) T_\mathrm{rad}(E)} .
\end{align}

We take $y\equiv \ah/a$ as the new independent variable instead of $t$. 
Substituting (\ref{eq:hardening_rate_xi}) into the above equation, we integrate it 
over $y$ with the initial condition $y=1, \xi(E,y=1)=1$ and obtain 
\begin{align}
\xi(E,y) &= \exp\left(-\frac{H_1 G\Mbin}{3\pi\eta L_\mathrm{circ}^2(E) T_\mathrm{rad}(E)} 
  h(y)\right) ,\\ 
h(y) &\equiv \int_1^y \frac{dy'}{y'\,S(y')} . \nonumber
\end{align}

Here $h(y)$ is another unknown function related to $S(y)$. Substituting the above expression 
back into (\ref{eq:hardening_rate_xi}), we can perform the integration over $E$, keeping $h$ 
as a fixed parameter, and obtain an equation of the kind $S(y) = g(h(y))$ with some function $g$. 
Then we express $h$ from the resulting equation as a function of $S$ and $y$, and finally 
differentiate $h$ by $y$ to obtain an ordinary differential equation for $S(y)$.

To illustrate this approach and acquire qualitative insight, we consider the asymptotic behavior 
of the system at large $t$ for the case of a power-law density profile,
$\rho(r)\propto r^{-\gamma}$. 
Pure power-law profiles are unphysical in the sense that 
the total mass is infinite and the gravitational potential can take arbitrarily large positive 
values, but the contribution of stars at large radii to the hardening rate is negligible, 
so that all quantities of interest are finite. 
We choose to define the gravitational potential of the stellar cusp so that its value at the center 
is $\Phi_0=0$; since we only consider stars that are not bound to the binary, their energies 
lie in the range $0\le E <\infty$.
Moreover, we may neglect the potential of the SBHs, because at late times most of the surviving 
loss-region stars are located at far larger distances than $r_\mathrm{infl}$. 
Then all dynamical quantities have power-law dependence on energy:
$f(E)\propto E^{-(6-\gamma)/(4-2\gamma)}$, 
$L_\mathrm{circ}(E)\propto E^{(4-\gamma)/(4-2\gamma)}$, 
$T_\mathrm{rad}(E)\propto E^{\gamma/(4-2\gamma)}$. 
Carrying out the integration in (\ref{eq:hardening_rate_xi}), we obtain
$S(y) \propto (h(y)/\eta)^{-(2+\gamma)/(8-\gamma)}$, and ultimately get the asymptotic dependence 
of the hardening rate on $a$:
\begin{align}  \label{eq:hardening_rate_asympt_triax}
S(a) \propto \left[\eta/\ln(\ah/a)\right]^{\frac{2+\gamma}{6-2\gamma}} .
\end{align}

In other words, the hardening rate due to depletion of centrophilic orbits drops with time 
(or $a^{-1}$) very slowly, and has a moderate dependence on the fraction of chaotic orbits 
in the model $\eta$. 
This expression is also valid in the limiting case of a singular isothermal sphere ($\gamma=2$).

We now follow a similar argument for the axisymmetric case. Here we again assume that orbits 
with $L^2<\eta L_\mathrm{circ}^2(E)$ are chaotic, but only those with 
$L_z<L_\mathrm{LC}\equiv \sqrt{2G\Mbin a}$ can interact with the binary. 
The mass of such ``useful'' chaotic orbits in the case of a hard enough binary, when 
$L_\mathrm{LC}<\sqrt{\eta}L_\mathrm{circ}$, is given by
\begin{align}  \label{eq:Mch_axisym}
M_\mathrm{ch}\!=\!\int_{\Phi_0}^0\!\! dE\,4\pi^2\,T_\mathrm{rad}(E)\,
2\sqrt{\eta} L_\mathrm{circ}(E) L_\mathrm{LC}(a)\,\xi(E,t)f(E),
\end{align}
and the expression for their depletion rate, analogous to (\ref{eq:depletion_triax}), is
\begin{align}  \label{eq:depletion_axisym}
\frac{d\xi(E,t)}{dt} = -\xi(E,t) \frac{H_1}{6\pi} 
  \frac{\sqrt{2G\Mbin a(t)}}{2\sqrt{\eta} L_\mathrm{circ}(E) T_\mathrm{rad}(E)} .
\end{align}

The solution is obtained along the same lines as for the triaxial case, with the different 
definition of $h(y)\equiv \int dy/(\sqrt{y}\,S(y))$. 
We again consider the asymptotical evolution described by (\ref{eq:hardening_rate_xi}),
(\ref{eq:depletion_axisym}) for an idealized scale-free model and obtain
\begin{align}  \label{eq:hardening_rate_asympt_axisym}
S(a) \propto \left\{ \begin{array}{ll}
  (\eta a/\ah)^{\frac{2+\gamma}{4-2\gamma}} &,\;\gamma<2,\\
  \exp\left[-C\ah/(\eta a)\right] &,\;\gamma=2 \end{array} \right.
\end{align}

Unlike the triaxial case, the hardening rate drops quickly with decreasing $a$. 
Moreover, in realistic non-scale-free systems the mass of chaotic orbits (\ref{eq:Mch_axisym}) 
that are both not yet depleted ($\xi(E,t)\sim 1$) and able to interact with the binary 
($L_z<L_\mathrm{LC}(a)$) is finite, and after some time drops below the mass of the binary 
itself, which means that the further evolution virtually ceases. 

\begin{figure*}
\begin{center}\includegraphics{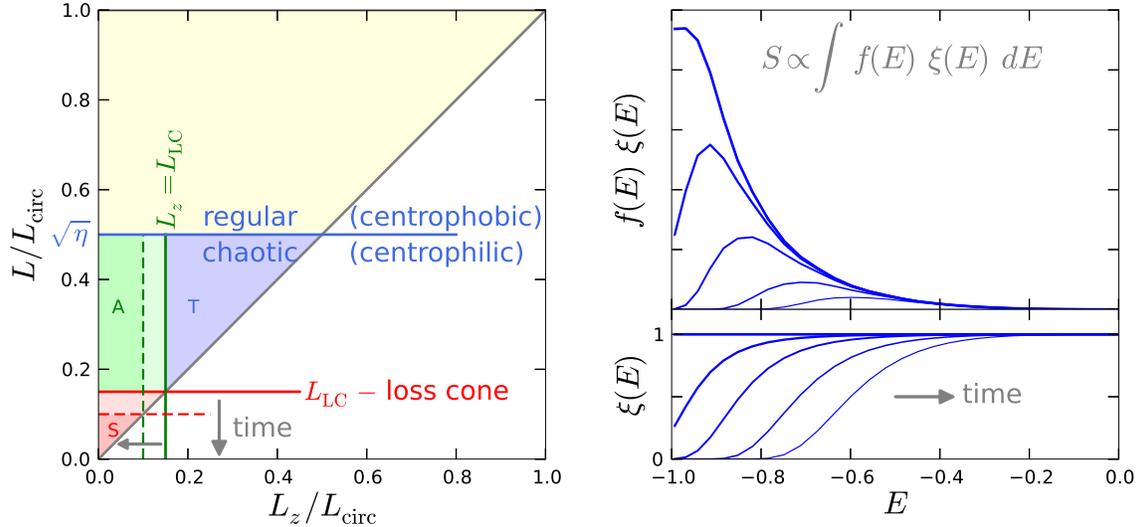}\end{center}\vspace{-0.5cm}
\caption{
Illustration of the loss region and hardening rate evolution in various geometries
(\textbf{S}pherical, \textbf{A}xisymmetric and \textbf{T}riaxial).\protect\\
Left panel shows the slice of the phase space ($L,L_z$) at a fixed energy $E$. 
The loss cone of the binary is the region $L<L_\mathrm{LC}\equiv\sqrt{2G\Mbin a}$; 
stars in that region will interact and be ejected by the binary in one dynamical time. 
In non-spherical geometries, the extended loss region consists of stars with 
$L<\sqrt{\eta}L_\mathrm{circ}$ (more correctly, only a fraction of such stars are on 
chaotic orbits, but this does not qualitatively change the picture).
In a triaxial geometry, these stars can wander anywhere in this region (shaded in blue and green)
due to collisionless effects (non-spherical torques), and can eventually get into the loss cone 
proper (shaded in red). 
In the axisymmetric case they are restricted to lines of constant $L_z$ and can only wander 
in vertical direction, thus the loss region is a strip $L_z<L_\mathrm{LC},
L<\sqrt{\eta}L_\mathrm{circ}$ (shaded in green). 
The population of stars inside the loss region is gradually depleted, and as the binary shrinks, 
$L_\mathrm{LC}$ decreases. Dashed lines show the boundary of the loss cone and the axisymmetric 
loss region at a later moment of time; the volume of the loss region in the axisymmetric 
case also shrinks with the binary, and the number of stars in this region drops both due to 
its decreasing volume and decreasing fraction of surviving stars. \protect\\
Right panel illustrates the depletion of the loss region population as a function of energy 
and time. 
Bottom plot shows the fraction of surviving stars $\xi(E)$ at various moments of time. Initially 
$\xi=1$ and it depletes faster at high binding energies; thinner curves correspond to later times. 
Top panel shows $\xi(E)$ multiplied by the distribution function; the integral under this curve 
gives the hardening rate at any given time, and initially is equal to the full-loss-cone hardening 
rate ($S_\mathrm{iso}$). Thus it is natural that the hardening rate is always smaller than 
$S_\mathrm{iso}$ and decreases with time.
}  \label{fig:phaseplane_dfevol}
\end{figure*}

\begin{table}
\caption{Long-term evolution of hardening rate in triaxial models.}  \label{tab:S_longterm}
First column is the model type, second and third are the full-loss-cone hardening rate 
and the radius of hard binary, in model units; 
last two columns are the dimensionless parameters of Equation~\ref{eq:S_powerlaw}, computed 
as best-fit values to the hardening rate found in collisionless Monte Carlo simulations.
\begin{tabular}{lllll}
\hline
model & $S_\mathrm{infl}$ & $\ah$ & $\mu$ & $\nu$ \\
\hline
default ($\gamma=1$, $q=1$, \\
$\;\;x\!:\!y\!:\!z=1\!:\!0.9\!:\!0.8$) & 20 & 0.0125 & 0.38 & 0.32 \\
$q=1/9$ & 30 & 0.004 & 0.33 & 0.32 \\
$x\!:\!y\!:\!z=1\!:\!0.8\!:\!0.6$ & 20 & 0.0125 & 0.96 & 0.51 \\
$\gamma=2$ & 1700 & 0.0022 & 0.25 & 0.62 \\
merger ($\gamma=1$), N=128k & 15 & 0.017 & 0.21 & 0 \\
merger, N=256k & 15 & 0.017 & 0.43 & 0.18 \\
merger, N=512k & 15 & 0.017 & 0.45 & 0.25 \\
merger, N=1024k& 15 & 0.017 & 0.87 & 0.47 \\
\hline
\end{tabular}
\end{table}

These findings are beautifully illustrated by the results of collisionless 
Monte Carlo models in Figure~\ref{fig:longterm}, bottom panel: 
the hardening rate in the triaxial case drops very gently with $\ah/a$ and closely follows 
its asymptotic expression (\ref{eq:hardening_rate_asympt_triax}), while in the axisymmetric 
case it decays much faster, following (\ref{eq:hardening_rate_asympt_axisym}), 
and then drops nearly to zero when all chaotic orbits are depleted. 
Since our galaxy models are not scale-free and gradually lose the triaxiality, 
the hardening rate in the collisionless triaxial model is better described by a somewhat 
steeper dependence on $a$, namely $S\propto a^\nu$ with $\nu$ ranging from $\sim 1/3$ for 
mildly triaxial $\gamma=1$ models to $\gtrsim 1/2$ for more strongly triaxial models or for 
steeper cusps ($\gamma=2$), see Table~\ref{tab:S_longterm}. 
The fact that steeper cusps result in a stronger slowdown of hardening rate can be explained 
by a more rapidly declining dependence of the distribution function on energy: 
once the centrophilic orbits close to the binary are depleted, there are less available stars 
at larger distances in models with steeper density profiles.
Figure~\ref{fig:phaseplane_dfevol} illustrates the above arguments about the size of the loss 
region and its depletion.

\enlargethispage{\baselineskip}

Finally, we consider the effects of relaxation on the long-term behavior of the hardening rate.
From the above discussion, it is clear that it may only matter for spherical and axisymmetric 
systems, because in the triaxial case the draining is the main mechanism that keeps the loss cone 
filled. In general, the relaxation rate scales as $N_\star^{-1}$, but it does not trivially 
translate into the hardening rate because of several complications.
First of all, the steady-state flux of stars into the loss cone has a different dependence 
on the size of the loss cone $L_\mathrm{LC}$ and the relaxation rate in the limits of empty 
and full loss cone regimes: in the former case, at a fixed energy it scales as 
$[N_\star\ln(L_\mathrm{circ}/L_\mathrm{LC})]^{-1}$, and in the latter -- as 
$(L_\mathrm{LC}/L_\mathrm{circ})^2$. Integrated over all energies with an appropriate boundary 
condition at each energy, the hardening rate has a weaker than $N_\star^{-1}$ scaling, which 
furthermore depends on the evolutionary stage: as $a$ gets smaller, a larger fraction of 
total flux comes from the full-loss-cone region. 
Second, the flux of stars into the loss cone is actually higher than the steady-state models 
would predict, because initially there are large gradients in the phase space: stars with 
$L\lesssim L_\mathrm{LC}$ are absent while at larger $L$ their distribution is nearly unchanged.
Time-dependent models for the loss cone repopulation \citep{MilosMerritt2003b} predict a flux 
that is several times higher than the steady-state value at early times. 

Overall, we find that in the spherical case the hardening rate drops with $N_\star$ rather 
mildly at $N_\star\lesssim 2\times 10^5$, but for larger $N_\star$ approaches the asymptotic 
$N_\star^{-1}$ scaling. In $N$-body simulations, however, it does not drop quite as fast at 
large $N_\star$. One possible reason might be the wandering (Brownian motion) of the binary 
\citep{QuinlanHernquist1997, ChatterjeeHL2003}, 
which effectively increases the size of the loss cone from $L_\mathrm{LC}^2=2G\Mbin a$ to 
$L_\mathrm{LC}^2=2G\Mbin r_\mathrm{wand}$. The wandering radius scales as 
$r_\mathrm{wand}\propto (m_\star/\Mbin)^{1/2}\propto N_\star^{-1/2}$ \citep[e.g.][]{Merritt2001}, 
thus for realistically large $N_\star$ it should remain below $a$ for the most part of evolution, 
and will not substantially increase the hardening rate.

Relaxation in the axisymmetric case occurs at the same rate as in the spherical system, but 
the effective size of the loss cone corresponds to the angular momentum of the chaotic region 
of the phase space $\sqrt{\eta}L_\mathrm{circ}$, which is then drained into the loss cone proper 
by the non-spherical torques and not by relaxation. This suggests that in the empty-loss-cone 
regime the steady-state flux is moderately (a factor of few) larger than in the spherical case 
\citep{MagorrianTremaine1999, VasilievMerritt2013}, because it depends only logarithmically 
on the effective size of the loss region. 
Time-dependent flux is again higher than the steady-state value, by larger factors than in 
spherical systems \citep[figure~13, left panel]{VasilievMerritt2013}. Given these uncertain 
complications, it is hard to derive more quantitative theoretical estimates for axisymmetric 
systems. The results of Monte Carlo simulations suggest that the hardening rate drops at least 
as $N_\star^{-1/2}$ in the range $10^6\le N_\star\le 10^9$, and probably even steeper at larger 
$N_\star$, which means that it is too slow for realistic galaxies to bring the binary to 
GW-dominated regime in a reasonable time. 

\subsection{Estimates of the coalescence time}  \label{sec:coalescence}

Motivated by the above arguments, we write the stellar-dynamical hardening rate $S_\star$ 
in an evolving model as 
\begin{align}  \label{eq:S_powerlaw}
S_\star(a) = \mu\, S_\mathrm{infl}\, (a/\ah)^\nu ,
\end{align}
where the dimensionless coefficient $\mu \lesssim 1$ defines the initial value of $S_\star$ 
at the moment of hard binary formation, expressed in the units of ``full-loss-cone rate'' 
(\ref{eq:S_infl}), and the exponent $\nu$ describes its decay with $a$.
Theoretical models of the previous section and the results of Monte Carlo simulations suggest 
that in the collisionless case, $\mu=\mathcal{O}(1)$ and $\nu\simeq 0.3\div0.6$ in triaxial 
models (see Table~\ref{tab:S_longterm}), while in the presence of relaxation $\nu\simeq 0$ 
and $\mu \ll 1$, scaling roughly as $\mu\simeq (N_\star/10^5)^{-1}$ in the spherical and 
$\mu\simeq (N_\star/10^5)^{-1/2}$ in the axisymmetric cases.

As the binary hardens, GW emission becomes more and more effective. 
The instantaneous hardening rate due to GW is $S_\mathrm{GW} = 1/(aT_\mathrm{GW})$, 
where $T_\mathrm{GW}$ is given by Equation~\ref{eq:hardening_GW}.
We denote $a_\mathrm{GW}$ to be the value of $a$ at which $S_\mathrm{GW}=S_\star$.
Using the definitions of $\ah$ (\ref{eq:a_hard}) and 
$S_\mathrm{infl}$ (\ref{eq:S_infl}) with $\mathcal A=4$, we obtain a simple estimate of 
the coalescence time for the case $\nu=0$ (i.e.\ $S_\star=\mathrm{const}$):
\begin{align}  \label{eq:T_coal_approx}
T_\mathrm{coal} &\approx 1.6\times 10^8 \mbox{ yr}  \times
  \left(\frac{r_\mathrm{infl}}{30\mbox{ pc}}\right)^2
  \left(\frac{\Mbin}{10^8\,M_\odot}\right)^{-1} \\
  &\times \mu^{-4/5}\, \left(\frac{4q}{(1+q)^2}\right)^{-1/5} f(e)^{1/5} . \nonumber
\end{align}

Let us now consider a qualitative model for the evolution of the binary driven by both 
stellar-dynamical and GW hardening.
\begin{align}  \label{eq:hardening_combined}
\frac{d(1/a)}{dt} &= S_\star(a) + S_\mathrm{GW}(a,e)  \\
  &= S_{\star,\mathrm{h}} \left(\frac{a}{\ah}\right)^\nu + 
  S_\mathrm{GW,h} \left(\frac{\ah}{a}\right)^5 \frac{f(e)}{f(e_\mathrm{h})} , \nonumber
\end{align}
where the values with subscript ``h'' refer to the moment of hard binary formation, 
and the eccentricity dependence $f(e)$ is given by Equation~\ref{eq:f_ecc}.
We further define a dimensionless parameter 
$\epsilon_\mathrm{GW}\equiv S_\mathrm{GW,h}/S_{\star,\mathrm{h}}$.
In all realistic situations, $\epsilon_\mathrm{GW} \ll 1$, meaning that at the early stage 
of evolution the hardening is driven by stellar encounters and not by GW emission.
If we assume that the eccentricity remains constant throughout the evolution, 
then the above equation yields the following time to coalescence
\begin{align}  \label{eq:T_coal_analytic}
T_\mathrm{coal} &= \frac{1}{S_{\star,\mathrm{h}} \ah} 
  \int_1^\infty \frac{dy}{y^{-\nu} + \epsilon_\mathrm{GW} y^5}  \\
  &= \frac{1}{4\epsilon_\mathrm{GW}\,S_{\star,\mathrm{h}} \ah} \:
  {}_2F_1\left( 1, \frac{4}{5+\nu};\; \frac{9+\nu}{5+\nu};\; 
  -\frac{1}{\epsilon_\mathrm{GW}} \right) .\nonumber
\end{align}
Here ${}_2F_1$ is the Gauss' hypergeometric function and $y \equiv \ah/a$.
In the limit $\epsilon_\mathrm{GW} \to 0$, the asymptotic value is 
\begin{align}
T_\mathrm{coal} &\simeq 1.7\times 10^8 \mbox{ yr}  \times
  \left(\frac{r_\mathrm{infl}}{30\mbox{ pc}}\right)^{\frac{10+4\nu}{5+\nu}}
  \left(\frac{\Mbin}{10^8\,M_\odot}\right)^{-\frac{5+3\nu}{5+\nu}}  \nonumber\\
  &\times \mu^{-\frac{4}{5+\nu}}\, \left(\frac{4q}{(1+q)^2}\right)^{\frac{3\nu-1}{5+\nu}}
  \!f(e)^{\frac{1+\nu}{5+\nu}}\; 20^\nu.  \label{eq:T_coal_asympt}
\end{align}

The ratio between $\ah$ and $a_\mathrm{GW}$, which describes how much the binary must shrink 
by stellar-dynamical processes before the GW emission takes over, is
\begin{align}
\frac{\ah}{a_\mathrm{GW}} &\simeq 160 \times 
  \left(\frac{r_\mathrm{infl}}{30\mbox{ pc}}\right)^{\frac{5}{10+2\nu}}
  \left(\frac{\Mbin}{10^8\,M_\odot}\right)^{-\frac{5}{10+2\nu}}  \nonumber\\
  &\times [\mu\,f(e)]^{\frac{1}{5+\nu}}\, \left(\frac{4q}{(1+q)^2}\right)^{\frac{4}{5+\nu}}
  0.4^\nu.  \label{eq:ah_over_agw}
\end{align}

The assumption of a constant eccentricity is not quite correct, though, as it both 
increases by stellar encounters, as described by Equation~\ref{eq:K_ecc}, and decreases 
due to GW emission, as given by Equation~\ref{eq:ecc_GW}. In the final stage of evolution, 
when one may neglect $S_\star$ compared to $S_\mathrm{GW}$, the following quantity is conserved:
\begin{align}  \label{eq:ecc_GW_integral}
a^{-1}\,e^{12/19}\,(304+121e^2)^{870/2299}\,(1-e^2)^{-1} = \mathrm{const} .
\end{align}

\begin{figure}
\includegraphics{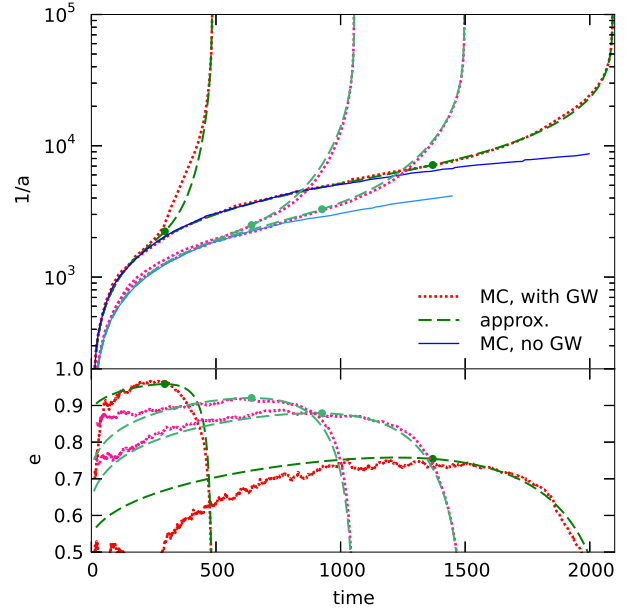}
\caption{
Evolution of the inverse semimajor axis (top panel) and eccentricity (bottom panel) for several 
triaxial models with no relaxation and the speed of light equal to 800 $N$-body velocity units.
Red dotted line is the Monte Carlo simulation, green dashed line is the result of numerical 
integration of equations (\ref{eq:ecc_GW}),(\ref{eq:K_ecc}),(\ref{eq:hardening_combined}), 
and blue solid line is the reference Monte Carlo simulation without GW emission. 
Green dot marks the transition to GW-dominated regime ($S_\star=S_\mathrm{GW}$).
The two evolutionary tracks ending at $t\simeq 1000$ and $t\simeq 1500$ are for an equal-mass 
system ($q=1$), and the other two are for mass ratio $q=1/9$.
} \label{fig:gw_evol_track}
\end{figure}

\begin{figure}
\includegraphics{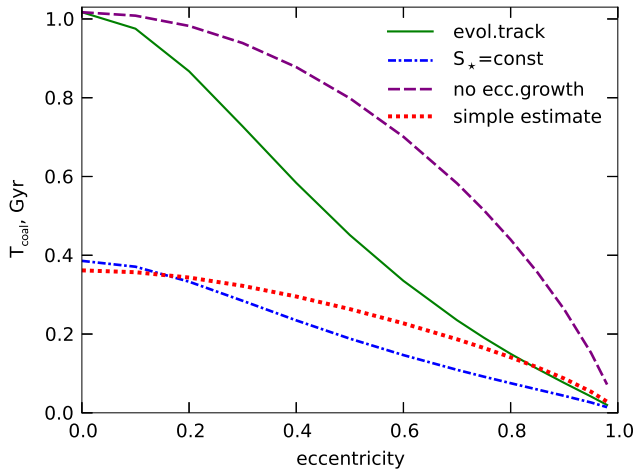}
\caption{
Coalescence time for triaxial models with $\Mbin=10^8\,M_\odot$ and $r_\mathrm{infl}=30$~pc, 
as a function of initial eccentricity. The initial hardening rate $S_\star$ is defined by 
(\ref{eq:S_powerlaw}) with $\mu=0.4$ and $\nu=1/3$, in accordance with Monte Carlo simulations 
for an equal-mass binary.
Green solid line is the result of numerical integration of the evolutionary track under 
our standard assumptions, purple dashed line is the evolutionary track computed without 
stellar-dynamical eccentricity growth (setting $A=0$ in Equation~\ref{eq:K_ecc}), 
blue dot-dashed line is the track computed for a constant $S_\star$ (i.e.\ $\nu=0$),
and red dotted line is the simple estimate (\ref{eq:T_coal_approx}) which also assumes 
a constant hardening rate and neglects the changes in eccentricity.
} \label{fig:T_coal}
\end{figure}

Evolutionary tracks may be computed by numerical integration of the coupled system of equations 
(\ref{eq:ecc_GW},\ref{eq:K_ecc},\ref{eq:hardening_combined}) for $a(t),e(t)$.
Figure~\ref{fig:gw_evol_track} shows several examples of evolutionary tracks computed using 
these equations, together with the results of Monte Carlo simulations of triaxial models 
with different initial eccentricity and mass ratio.
We used the initial eccentricity as a free parameter in the evolutionary tracks, and adjusted 
to match the coalescence times from the Monte Carlo simulations: since the eccentricity varies 
rather erratically at early stages of evolution, it is hard to match these two curves without any
tuning, but with this one free parameter one gets a quite good agreement at late stages of evolution.

Figure~\ref{fig:T_coal} shows the coalescence times computed for triaxial models with 
$\Mbin=10^8\,M_\odot$, using the simple estimate (\ref{eq:T_coal_approx}) and numerically 
computed evolutionary tracks, as a function of initial eccentricity. 
For $e=0$, the coalescence time $T_\mathrm{coal}^{e=0}$ can be obtained analytically 
(Equation~\ref{eq:T_coal_asympt}).
For arbitrary eccentricity, it may be approximated as 
\begin{align}  \label{eq:T_coal_triax_ecc}
T_\mathrm{coal} &\approx T_\mathrm{coal}^{e=0} \times (1-e^2) \left[ k + (1-k)(1-e^2)^4 \right] ,\\
k &= 0.4 + 0.1\,\log_{10}(\Mbin/10^8\,M_\odot) . \nonumber
\end{align}

It is instructive to compare the above expression to the simple estimate (\ref{eq:T_coal_approx}), 
which predicts $T_\mathrm{coal}\propto (1-e^2)^{7/10}$ for the case of constant hardening rate 
and no evolution of eccentricity. For $e=0$ the latter underestimates the coalescence time 
by a factor $\sim 3$, but at higher eccentricities the detailed evolutionary tracks come closer 
to the simple estimate, because the slowdown of stellar-dynamical hardening is compensated by 
the increase in eccentricity at the same evolutionary stage. 
The rather weak trend of the eccentricity-dependent factor with $\Mbin$ reflects the larger 
ratio between $\ah$ and $a_\mathrm{GW}$ for smaller $\Mbin$ (Equation~\ref{eq:ah_over_agw}), 
thus for them the stellar-dynamical increase in $e$ is larger and consequently brings GW 
on stage earlier.

The coalescence time very weakly depends on $\Mbin$ -- if we adopt the $M_\bullet-\sigma$ relation 
in the form (\ref{eq:Msigma}), then $T_\mathrm{coal}\propto \Mbin^{0\div0.1}$. It also only moderately 
depends on the initial eccentricity: despite the much steeper dependence of the GW hardening rate 
on $e$, most of the time is spent on the stellar-dynamical hardening stage. As a consequence,
the efficiency of stellar-dynamical hardening $\mu$ is as important as the eccentricity. 
In the triaxial case, the efficiency is high enough and it decays slowly enough that 
for all reasonable parameters the coalescence time is shorter than the Hubble time.
It also rather mildly depends on the fraction of chaotic orbits $\eta$, which itself is determined 
by triaxiality; the caveat is that the latter changes in the course of evolution, but generally 
even a slight triaxiality is enough to support the required population of centrophilic orbits.

By contrast, in the collisionless axisymmetric case, the hardening rate slows down so rapidly 
that the binary stalls at a separation too large for efficient GW emission, unless 
the eccentricity is very high. 
If we take into account relaxation-driven repopulation of the loss cone,
then Equation~\ref{eq:T_coal_approx} suggests that the coalescence time may be shorter than 
the Hubble time for $\Mbin \lesssim 10^8\,M_\odot$ and moderate eccentricity 
(under the optimistic assumption that the hardening rate, determined from Monte Carlo simulations 
to be proportional to $N_\star^{-1/2}$, stays at this relatively high value for a much longer 
time than these simulations were run). 
But in a realistic galaxy, even relatively minor perturbations from axisymmetry would create 
a sufficient reservoir of centrophilic orbits, whose draining maintains a much higher hardening 
rate than the relaxation can provide.

\section{Evolution of merger remnants}  \label{sec:evol_merger}

The isolated models considered above serve as a controlled experiment that helps to understand 
the physical mechanisms responsible for the joint evolution of stars and the binary. 
However, the initial conditions were somewhat artificial in that we have set up initial models 
in almost perfect equilibrium. 
In the cosmological setting, binary SBHs are expected to form via galaxy mergers, and the merger 
remnants could well have a complex and evolving structure quite unlike our idealized models.

\begin{figure}
\includegraphics{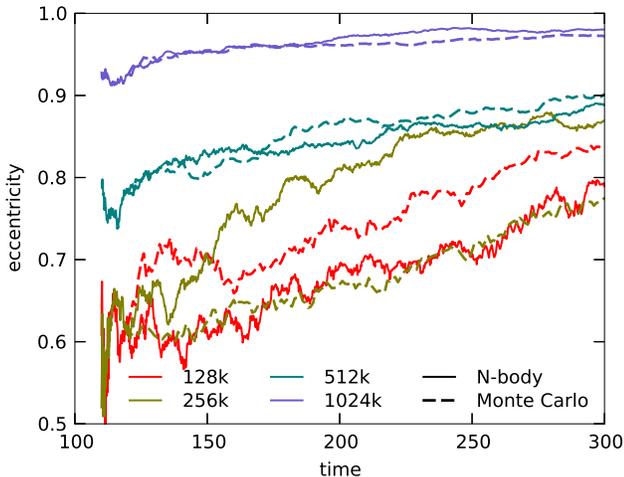}
\caption{
Binary ecccentricity in merger simulations: solid lines are $N$-body and dashed lines are 
Monte Carlo models with relaxation; from bottom to top: $N=\{128,256,512,1024\}\times10^3$.
}  \label{fig:merger_eccentricity}
\end{figure}

In this section we consider a limited set of merger simulations, similar to those of 
\citet{KhanJM2011}. We set up two identical spherical $\gamma=1$ Dehnen models with scale radius 
and mass equal to unity, each containing a central SBH with mass $M_\bullet=0.01$.
They are put on an elliptical orbit with initial separation 20 and relative velocity 0.1; 
the first encounter between the galaxies occurs at $t\simeq 80$, the second at $t\simeq 100$, 
and by $t=110$ the two nuclei are well merged and the binary is formed. 
The spherically averaged density profile of the central region of the merger remnant 
right after hard binary formation is well described by a $\gamma=0.5$ Dehnen profile 
with unit scale radius and total mass $\simeq1.7$. 
We evolve the system until $t=300$, using the same direct-summation code $\phi$GRAPEch, and 
extract a snapshot at a time $t\simeq 120$ when the binary semimajor axis $a=0.01\lesssim \ah$; 
this snapshot then constitutes the initial conditions for the Monte Carlo simulations.
We ran four simulations with particle numbers $N=\{128,256,512,1024\}\times10^3$.
The morphology of the merger remnants was roughly the same, but the initial eccentricity of 
the binary at the moment of its formation was systematically larger for higher-resolution 
simulations (Figure~\ref{fig:merger_eccentricity}), even though the orbit parameters of the 
merging galaxies were identical; thus these four Monte Carlo models are not exactly equivalent.

Unlike the isolated models, merger remnants do not exhibit genuine triaxial symmetry,
but only a reflection symmetry (i.e.\ even after a rotation that aligns the major axis with the $x$-coordinate axis, 
they are not invariant with respect to a flip about either the $x$ or $y$ axes, but stay the same under 
a simultaneous inversion of both axes -- like a barred spiral galaxy in which the spiral arms break 
the triaxial symmetry of the bar).
We therefore kept all reflection-symmetric (even $l$ and all $m$) terms in the spherical-harmonic 
expansion of the potential in the Monte Carlo code. 
The density profile rotates with a non-uniform angular speed, which precludes the use of 
a rotating reference frame; since the rotation is rather slow, we just updated the expansion 
coefficients frequently enough (each 1 time unit) to track this figure rotation.
We only used terms up to quadrupole ($l=2$), therefore smoothing out all irregularities and 
small-scale structure that might be present in the merger remnant, and retaining only the global 
non-spherical features. We have checked that using higher-order harmonics does not substantially 
change the results. In addition, we have run simulations with imposed axisymmetry (setting all
$m\ne0$ terms to zero). For each of the four values of $N$, we have performed Monte Carlo 
simulations with relaxation (corresponding to $N_\star=N$ in the actual $N$-body system) 
and without relaxation.

\begin{figure}
\includegraphics{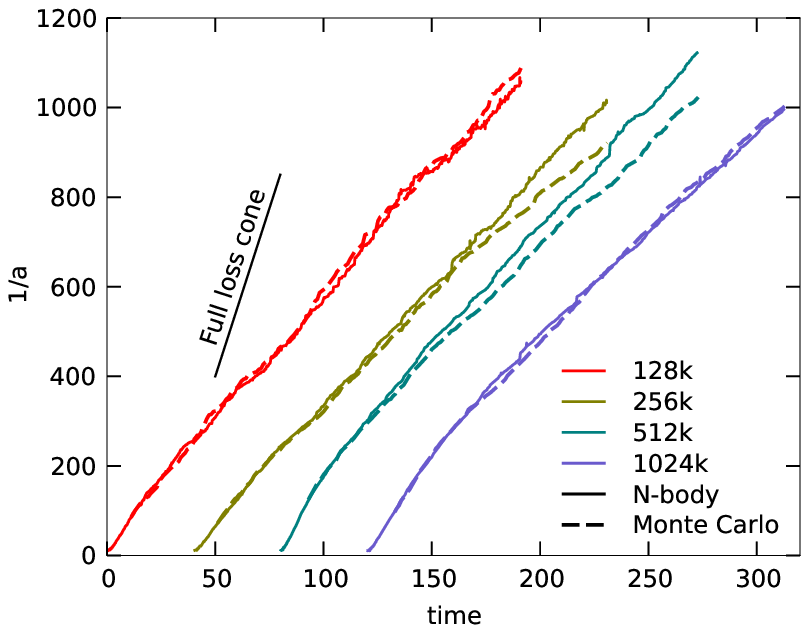}
\includegraphics{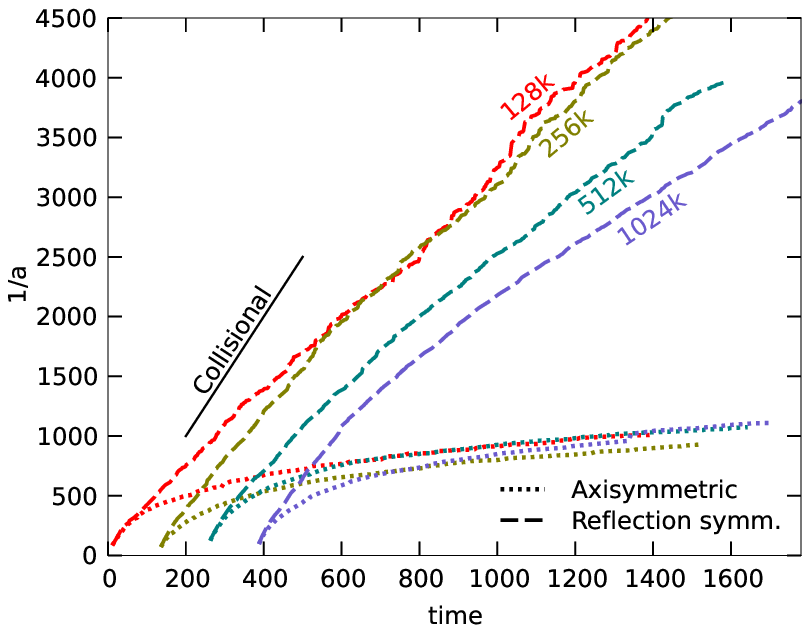}
\caption{
Evolution of inverse semimajor axis of the binary in merger simulations. 
Top panel compares $N$-body (solid lines) and Monte Carlo models with relaxation 
(dashed lines), from left to right: $N=\{128,256,512,1024\}\times10^3$ (curves are shifted 
horizontally for clarity). 
Bottom panel shows the long-term evolution of collisionless Monte Carlo models, with imposed 
axisymmetry (dotted lines) and with only reflection symmetry (i.e., nearly triaxial models 
with figure rotation, dashed lines). Solid line shows the average hardening rate in collisional 
simulations. 
}  \label{fig:merger_hardening}
\end{figure}

The results, shown in Figure~\ref{fig:merger_hardening}, can be summarized as follows. 
In the $N$-body simulations, the hardening rate was found to be almost independent of $N$, 
in agreement with other studies. However, the highest-$N$ model had a slightly lower hardening 
rate at late times. 
The weaker $N$-dependence of hardening rate in merger remnants compared to idealized isolated 
models could have a number of reasons, e.g.\ perturbations from the decaying large-scale clumps 
and inhomogeneities in the merger remnant are presumably independent of $N$ and add up to 
the conventional two-body relaxation. This conjecture is supported by the fact that 
the rate of energy and angular momentum diffusion measured in the simulations is almost 
independent of $N$ and much higher than the rate expected from two-body relaxation.
The eccentricity, being quite high at the formation time, increased further by the end of 
the simulation, up to $\gtrsim 0.98$ for $N=10^6$. 
We note that the measured hardening rate was $\sim 3$ times lower than $S_\mathrm{full}$, again 
underlining that even in strongly asymmetric and dynamic systems the loss cone is at least 
partially depleted at the highest binding energies. 

While the results Monte Carlo simulations in this section should be regarded as preliminary, 
it is remarkable that the agreement in hardening rates between $N$-body and Monte Carlo 
simulations that included relaxation is reasonably good (Figure~\ref{fig:merger_hardening}, 
top panel). 
Relaxation driven by large-scale fluctuations in the merger remnant might be underestimated 
in the Monte Carlo method because of two factors: 
(a) the potential represented by spherical-harmonic expansion smooths out small-scale clumps, 
(b) the interval between potential updates is short enough to track global changes in 
the potential, but may not represent higher-frequency transient perturbations.
Thus we should expect somewhat lower hardening rate in Monte Carlo simulations compared to 
$N$-body simulations; in fact it turned out to be comparable and even sometimes higher. 
We defer a more detailed study of Monte Carlo models of merger remnants for a future work.

In the collisionless regime, the hardening rate is initially only slightly lower than in 
the $N=10^6$ simulation with relaxation. However, it slows down later on, much like in the case 
of isolated triaxial systems (Table~\ref{tab:S_longterm}). 
Again, this is due to two factors: depletion of centrophilic orbits and gradual decrease of 
triaxiality. The shape of the merger remant is moderately flattened ($z/x\simeq 0.75$ at all 
radii throughout the simulation), and the triaxiality is quite subtle: $y/x \simeq 0.9$ right 
after merger and increases to $\gtrsim 0.97$ toward the end of simulation. 
Nevertheless, this appears to be enough to sustain the reservoir of centrophilic orbits needed 
to keep the binary shrinking. 
Even with zero eccentricity, the binary would merge in $\sim 0.5$~Gyr; with such high eccentricity 
as in our simulations, this time would be an order of magnitude shorter. 
If we force the potential to be axisymmetric, the evolution slows down much faster
(Figure~\ref{fig:merger_hardening}, bottom panel); by the end of simulation $a\simeq 0.05\ah$, 
which is not enough to ensure merger in a Hubble time unless the eccentricity is $\gtrsim 0.8$.

\section{Discussion and conclusions}  \label{sec:discussion}
\subsection{Summary of our results}  \label{sec:summary}

In the present work, we have considered a stellar-dynamical solution to the final-parsec problem 
-- the evolution of binary SBHs driven by encounters with stars in a galactic nucleus, 
as its orbit shrinks from the radius of a hard binary ($\ah \sim 1$~pc) to the separation 
$a_\mathrm{GW}\sim 10^{-2}\,\ah$ at which GW emission becomes effective. 
The central difficulty is to ensure a continuous supply of stars onto low-angular-momentum 
orbits, where they can be scattered by the binary and carry away its energy and angular 
momentum. This region of phase space, dubbed the loss cone, is quickly depleted by the binary 
once it becomes hard, so an efficient refilling mechanism is required to enable continued hardening. 
In a perfectly spherical galaxy, this can only be achieved by two-body relaxation, which is 
not sufficient to bring the binary to coalescence in a Hubble time, except in the smallest 
galaxies \citep{MerrittMS2007} -- hence the problem. 
We have focused on the additional mechanisms of loss-cone repopulation that exist in 
non-spherical (axisymmetric and triaxial) galaxies.

We addressed this problem using a variety of methods, but primarily with \textsc{Raga},%
\footnote{The code is available at \url{http://td.lpi.ru/\~eugvas/raga/}}
a novel stellar-dynamical Monte Carlo code that is able to follow the evolution of non-spherical 
stellar systems under the influence of two-body relaxation, the magnitude of which can be 
adjusted -- unlike conventional $N$-body simulations in which the relaxation is essentially 
determined by the number of particles. We extended the code to include interactions between 
stars and the massive binary, by following the trajectories of particles in the superposition 
of the smooth galactic potential plus the time-dependent potential of the two SBHs as they 
orbited one another, and used conservation laws to determine the reciprocal changes in 
the binary orbital parameters. 

We used a large suite of direct-summation $N$-body simulations to verify that the Monte Carlo 
code accurately describes the evolution of the binary for a wide range of parameters -- 
mass ratio, eccentricity, number of particles and the shape of the galaxy model. 
Then we extended the Monte Carlo simulations into the range of $N_\star$ much larger than is 
presently accessible for conventional $N$-body simulations, including the intriguing 
collisionless limit ($N_\star=\infty$), which is nearly achieved in real galaxies. 
We determined the scaling laws and asymptotic behavior of the stellar-dynamical hardening 
rate from rather simple analytic arguments, and confirmed these findings with the Monte Carlo 
simulations. Taking into account GW emission, the evolution of binary semimajor axis $a$ and 
eccentricity $e$ can be described by a simple system of differential equations, which we used to 
determine the coalescence time as a function of initial parameters of the binary and the galaxy; 
these evolutionary tracks were again verified by Monte Carlo simulations for a few test cases. 

Our main results can be summarized as follows:
\begin{enumerate}
  \item The binary continues to shrink as long as there are stars in the so-called loss region -- 
the region of phase space from which stars can precess into the loss cone proper 
(that is, $L^2<L_\mathrm{LC}^2\equiv 2G\Mbin a$) due to non-spherical torques. 
In the spherical case, the loss cone and loss region are the same. 
In the triaxial case, the loss region consists of mostly chaotic orbits, whose total mass is 
roughly proportional to the mass of galaxy with a coefficient $\eta\ll 1$ determined by the 
degree of flattening and triaxiality, and unless the galaxy is nearly axisymmetric, the mass 
of stars in the loss region is much larger than $\Mbin$.
In the axisymmetric case, the volume of the loss region shrinks with $a$ as $\sqrt{\eta a}$, 
halfway between the spherical and triaxial cases. 
  \item The stellar-dynamical hardening rate $S_\star$ is always smaller than the value 
$S_\mathrm{full}$ corresponding to a full loss cone. 
This is explained by a gradual depletion (draining) of the loss region, which occurs faster 
at high binding energies. As a rough estimate, decreasing $a$ by a factor of two requires 
ejection of stars with total mass $\sim \Mbin$.
In the collisionless limit, the hardening rate declines with $a$ very slowly for a triaxial system, 
because the total mass of loss region stars is typically large compared to $\Mbin$.
By contrast, in an axisymmetric system, the volume of the loss region also shrinks with $a$, 
and it depletes much faster; thus the hardening rate declines rapidly and drops nearly to zero 
at $a\lesssim 0.1\ah$. This is not enough to bridge the gap to the GW-dominated regime.
  \item Taking into account relaxation-driven repopulation of the loss region does not change 
the results in the triaxial case very much, because the hardening is dominated by draining of 
stars that are initially in the loss region, but accounting for relaxation does change 
the dynamics in the axisymmetric and spherical cases dramatically. 
After the initial population of the loss region is nearly depleted, it is refilled 
by two-body relaxation, maintaing the hardening rate at a nearly constant level that depends 
on $N_\star$. 
The axisymmetric case offers a didactic example of how much a system with $N_\star\sim10^6$, 
typical of present-day, high-fidelity $N$-body simulations, 
or even $N_\star\sim10^8$, can differ from the collisionless limit in its long-term evolution.
  \item Coalescence times estimated for triaxial galaxies weakly depend on the mass of 
the binary, its mass ratio, or the degree of triaxiality (provided that the departure from 
axisymmetry is larger than a few percent). 
Coalescence times fall in the range from a few Gyr for almost circular binaries, to 
$\lesssim 10^8$~yr for very eccentric ones.
For a given combination of the binary mass $\Mbin$ and its radius of influence $r_\mathrm{infl}$,
the coalescence time can be computed using Equation~\ref{eq:T_coal_asympt} for a circular orbit, 
and using Equation~\ref{eq:T_coal_triax_ecc} for an eccentric orbit, 
where the typical values of the dimensionless parameters $\mu\sim 0.2\div1$ and $\nu\sim 0\div0.5$ 
are listed in Table~\ref{tab:S_longterm}.
This time is up to a factor of few times greater than the simple estimate (\ref{eq:T_coal_approx}) that 
does not account for the decrease of the hardening rate with time.
\end{enumerate}

\subsection{Comparison with previous work}  \label{sec:comparison}

Consider first the spherical case, which has been extensively studied by $N$-body 
simulations \citep[e.g.][]{QuinlanHernquist1997,MilosMerritt2001,HemsendorfSS2002,BerczikMS2005,
MerrittMS2007}, methods based on scattering experiments \citep[e.g.]{Quinlan1996,Sesana2010,
MeironLaor2012}, or Fokker--Planck models \citep{MilosMerritt2003b,MerrittMS2007}.
It is generally accepted that in the spherical case the binary quickly depletes the loss cone 
and its evolution nearly stalls at a value of $a$ just a few times smaller than $\ah$. 
A number of effects may moderately decrease the stalling radius or increase the relaxation rate.
The secondary slingshot -- the re-ejection of stars that have once  interacted with the binary 
but did not gain enough energy to escape the galaxy -- leads to a gradual ($\propto \log t$) 
increase of $1/a$ at late times \citep{MilosMerritt2003b}, and this trend was indeed found 
in our collisionless Monte Carlo simulations. Once these stars are completely eliminated, 
evolution of the binary finally stalls in the purely collisionless case; 
\citet{Merritt2006} and \citet{SesanaHM2007} found that the stalling radius is only a few times 
smaller than $\ah$ over a wide range of binary mass ratios and cusp density profiles.
Similarly, time-dependent solution of the Fokker--Planck equation describing the relaxation of 
stars in angular momentum yields a higher rate of loss-cone repopulation at early times than 
the steady-state flux, due to sharper gradients in the phase space \citep{MilosMerritt2003b};
this is taken into account automatically in the Monte Carlo scheme, and does not substantially 
affect the overall evolution. 

Brownian motion is not accounted for in our method, and this may be the reason for 
the discrepancy between Monte Carlo and $N$-body hardening rates at high $N$ in spherical systems.
\citet{QuinlanHernquist1997} and \citet{ChatterjeeHL2003} argued that wandering of the binary 
may explain the very weak dependence of hardening rate on $N$ found in their simulations. 
However, as summarized by \citet{MakinoFunato2004}, several factors complicate the interpretation 
of their result: the $N$-body algorithm  used in that paper is less collisional than traditional 
$N$-body schemes, but not entirely free of relaxation, and the unequal masses of particles 
mean that the granularity of the potential depends on the evolutionary stage. 
The amplitude of Brownian motion of the binary, while larger than for a single SBH 
of the same mass \citep{Merritt2001}, scales roughly as $N^{-1/2}$, and in real galaxies 
would be smaller than the size of the loss cone for most part of the evolution. 
Using somewhat different arguments, \citet{MilosMerritt2003b} estimated the timescale 
of loss-cone refilling by Brownian motion and concluded that this effect is unlikely 
to substantially affect the evolution.

Consider next the case of non-spherical galaxies, which seem to offer a more promising way 
to solve the final-parsec problem via collisionless dynamics.
\citet{Yu2002} estimated draining rates for the loss regions in axisymmetric and triaxial 
galaxies, using arguments similar to those in our section~\ref{sec:evol_longterm_theory}. 
\citet{Yu2002} concluded that the loss region in triaxial galaxies is not likely to be depleted 
if the flattening parameter, responsible for the fraction of centrophilic orbits, is 
$\gtrsim 0.05$, although the plots in which she showed evolution timescales demonstrate 
the effect of gradual depletion of the loss region only for the case of small flattening.
For axisymmetric systems, \citet{Yu2002} estimated that the loss region \citep[or ``loss wedge'', 
in the terminology of][]{MagorrianTremaine1999} can be depleted rather quickly in many cases, 
and the relaxation-limited evolution timescales are still longer than the Hubble time. 
Thus our conclusions qualitatively agree with that study.

\citet{MerrittPoon2004} considered self-consistent triaxial models of galactic nuclei with 
SBHs which had a significant fraction of centrophilic orbits. They solved the evolutionary 
equations similar to (\ref{eq:hardening_rate_xi},\ref{eq:depletion_triax}), and found that 
the hardening rate is nearly independent of time, but scales with the square of the fraction 
of chaotic orbits. 
Analysis of the data plotted in their figure~6 suggests that $1/a\propto t^{0.7\div 0.9}$, 
more in line with our findings in the present study and in \citetalias{VasilievAM2014}. 
For the case of a  singular isothermal cusp ($\gamma=2$), 
Equation~(\ref{eq:hardening_rate_asympt_triax}) suggests that the asymptotic hardening rate 
scales as the squared fraction of chaotic orbits $\eta$, in agreement with their results. 
Their equation~55 implies a hardening rate $\simeq \eta^2 S_\mathrm{full}$, 
and our Monte Carlo simulation of a $\gamma=2$ Dehnen model has on average a similar hardening 
rate with $\eta=0.2$, even though it declines with time. 
Thus the basic conclusion of that paper, that even a moderate amount of triaxiality is 
sufficient to drive the binary to coalescence in less than a Hubble time, is corroborated by 
our simulations and asymptotic analysis, even though some details differ (most importantly, 
their neglect of the slowdown of the hardening rate). 

Most other studies to date have assumed or inferred that the loss cone must be kept nearly 
full in non-spherical systems, using various arguments.
\citet{HolleySigurdsson2006} explored the properties of orbits in a triaxial galaxy with 
a central SBH. They estimated that the time required to change the angular momenta of particles 
due to collisionless torques was much shorter than the Hubble time, and concluded that the loss 
cone must remain full. However, this argument does not take into account the draining of 
the loss region, and they did not analyze the evolution of their model under this process.
More recently, \citet{LiHK2014} performed a similar analysis for a nearly axisymmetric model, 
which in fact was slightly triaxial in the central part. They computed the mass of stars 
belonging to orbits that are able to come into the loss cone with a size corresponding 
to the initial stage of binary evolution, which was several times larger than $\Mbin$.
From this they concluded that the loss cone should remain well populated during the subsequent
evolution, but their estimate did not take into account that the volume of the loss region 
also shrinks along with the binary semimajor axis in the axisymmetric case. It is unclear 
whether the slight triaxiality of their model would be sufficient to support enough truly 
centrophilic orbits during the entire evolution.

\citet{Sesana2010} considered a hybrid model for binary evolution, based on the hardening 
and eccentricity growth rates computed from scattering experiments. He assumed that after 
the initial phase of formation of a hard binary, accompanied by the erosion of the stellar cusp 
at $r\lesssim r_\mathrm{infl}$, the subsequent evolution occurs in the full-loss-cone regime, 
i.e.\ the hardening rate is given by $S_\mathrm{uniform}$ (Equation \ref{eq:S_uniform})
with density and velocity dispersion computed at $r_\mathrm{infl}$. 
Thus his evolutionary tracks are similar to our calculations in Section~\ref{sec:coalescence} 
with parameters $\mu=1$ (the loss cone is full) and $\nu=0$ (the hardening rate does not decrease
with time). As Figure~\ref{fig:T_coal} and Equation~(\ref{eq:T_coal_approx}) show, in this case 
the coalescence time is shortened by a factor of few with respect to 
the more conservative assumptions proposed in our study.

More recently, \citet{SesanaKhan2015} compared the evolutionary tracks from the hybrid model 
of \citet{Sesana2010} with those obtained by $N$-body simulations of mergers \citep{Khan2012}.
They found reasonable agreement for the eccentricity evolution and hardening rate, 
even though the latter was somewhat lower and declined with time in $N$-body simulations. 
They ascribed this to the gradual decrease of the density and corresponding increase of 
$r_\mathrm{infl}$. As we have argued in Section~\ref{sec:evol_longterm_theory}, the decline of 
hardening rate is mostly caused by the depletion of centrophilic orbits
at all radii, rather than simply the decrease of density at the influence radius. 
For instance, in our collisionless simulations of isolated models with $\gamma=1\,(2)$, 
the actual hardening rate dropped by a factor of 5 (10) by the end of simulations, while 
the density at $r_\mathrm{infl}$, and correspondingly the full-loss-cone hardening rate 
$S_\mathrm{full}$, decreased by less than 30\%. 
The coalescence times quoted in \citet{SesanaKhan2015} are longer than ours
due to a different $\Mbin-r_\mathrm{infl}$ relation adopted in that paper.
We stress, however, that their estimates are based on the hardening rates from collisional 
$N$-body simulations, while coalescence times for collisionless systems, advocated in 
the present study, are up to a few times longer for the same galaxy parameters.

Studies based on $N$-body simulations have generally observed little or no dependence of 
the hardening rate on $N$ in non-spherical galaxy models that were formed via mergers \citep{PretoBBS2011,KhanJM2011,
Khan2012} or created as isolated models \citep{BerczikMSB2006,KhanHBJ2013}. 
This result has been interpreted as an indication that the loss cone remains nearly full, 
although \citet{BerczikMSB2006} reported that the hardening rate was gradually decreasing 
with time, suggesting that the reservoir of centrophilic orbits was being depopulated.
A similar trend can be seen in other merger simulations (e.g.\ \citealt{PretoBBS2011}, Fig.1, 
or \citealt{Khan2012}, Fig.2); in the latter paper, the models with steeper cusps displayed 
a systematically more rapid decline of hardening rate with time. 
All these trends are in agreement with our findings, even though the previous studies did not 
highlight them. 
We note, however, that in our simulations the hardening rate always turns out to be substantially 
lower than $S_\mathrm{full}$ for large enough $N$.
This might seem to be in contrast with other studies that report a hardening rate comparable
to $S_\mathrm{full}$ for non-spherical systems. However, a more detailed examination suggests 
that the apparent discrepancy can be attributed to different definitions of the full-loss-cone 
rate, and to different normalizations of the stellar density profile. For instance, in 
\citet{KhanHBJ2013} and \citet{HolleyKhan2015} the flattened models were created by adiabatic 
squeezing of the original density profile, and hence their scale radii are roughly a factor of two
smaller than ours, as can be seen in Figure~1 of the latter paper. From Equation~(\ref{eq:S_infl}),
it is  apparent that this translates to a hardening rate $\sim6$ times higher than ours.
We have re-simulated their models with our $N$-body code and found generally good agreement with 
their results.

Most importantly, as we have argued in \citetalias{VasilievAM2014} and this study, 
the hardening rates in $N$-body simulations are dominated by,
or at least have a significant contribution from, collisional effects even for $N\sim10^6$, 
thus it is not easy to extrapolate these results to real galaxies. 
Using the Monte Carlo method, we were able to reach the collisionless regime, which turned out 
to be very different for axisymmetric and triaxial galaxies, while in $N$-body simulations of 
\citetalias{VasilievAM2014} they looked nearly the same. 

\subsection{Single and binary black holes}  \label{sec:single_binary}

It is also instructive to compare  loss-cone theory in the single and binary SBH cases.
The first obvious difference is the much larger size of the loss cone in the case of a binary.
As a consequence, the relaxation-driven repopulation of the loss cone almost always occurs 
in the empty-loss-cone regime; on the other hand, collisionless changes in angular momentum 
due to non-spherical torques occur on the same dynamical timescale as the depletion of the loss 
cone, so that it always remains partially populated in non-spherical systems. 
A second important factor is that the size of the loss cone decreases as the binary shrinks, 
while in the case of a single SBH it can only grow. 
Third, the mass of stars needed to be delivered into the loss cone of the binary is a few times 
larger than the mass of (the lighter component of) the binary, while for single SBHs  
the accreted mass in stars is typically small compared to the SBH mass. 

These three factors explain the fundamental difference between collisionless spherical and 
axisymmetric systems, on the one hand, and triaxial ones, on the other hand. For the latter ones, 
the evolution is almost entirely driven by draining of centrophilic orbits, whose total mass 
is much larger than $M_\bullet$ and furthermore does not depend on the size of the loss cone. 
Interestingly, in the case of a single SBH most of captured stars arrive from regular pyramid 
orbits inside $r_\mathrm{infl}$, while in the case of the binary the loss region consists 
mainly of chaotic orbits outside $r_\mathrm{infl}$; this explains the slightly different time 
dependence of the draining rates.
In the axisymmetric case, the volume of the loss region composed of chaotic orbits that can be 
delivered into the loss cone by collisionless torques, shrinks along with the binary semimajor 
axis, and its orbit population is nearly depleted before the binary reaches the GW-dominated 
regime. The subsequent evolution is determined by the rate at which this loss region is 
repopulated by relaxation. Since the volume of this region is still much larger than the volume 
of the loss cone proper, it is more easily repopulated in axisymmetric than in spherical systems. 
The same is true for single SBHs; the fact that for them the difference between axisymmetric and 
triaxial systems is much less than between spherical and axisymmetric ones 
\citep[Figure~4]{Vasiliev2014a} stems largely from the adopted isotropic (non-depleted) initial 
conditions for the relaxation. On the other hand, in the case of a massive binary the phase 
space is already depleted out to much larger values of angular momentum than the current 
loss cone boundary, thus it takes longer for the relaxation to resupply the loss region. 

In short, loss cone theory in non-spherical systems is a delicate interplay between collisional 
and collisionless effects, and the outcome depends on the evolutionary history of the loss cone,
as well as the changes in the global structure of the system (its shape and phase-space gradients). 
Only using a combination of various approaches -- $N$-body simulations, orbit analysis, Monte Carlo 
methods and scaling arguments -- can one hope to understand the behavior of realistic stellar systems.

\subsection{Conclusions}  \label{sec:conclusions}

The evolution of binary SBHs in gas-poor galaxies is determined by the rate of slingshot 
interactions with stars in the loss cone -- the low-angular-momentum region of the phase space. 
The fact that the loss cone is quickly depleted in idealized spherical systems gave rise to 
the final-parsec problem. Repopulation of the loss cone occurs both due to collisional 
and collisionless effects; the latter are only relevant in non-spherical systems. 
We have developed a Monte Carlo method that can efficiently deal with both collisionless and 
collisional evolution, and used it to show that in the collisionless limit, the repopulation 
is efficient if the galaxy is even slightly triaxial. 
To the extent that mergers result in galactic shapes that are not exactly axisymmetric, 
our results imply that the final-parsec problem does not exist in most galaxies. 

\acknowledgements
We thank K.~Holley-Bockelmann and F.~Khan for making available to us data from
their $N$-body simulations, which were used in making the comparisons described in
Section \ref{sec:comparison}. 
E.V. thanks Alberto Sesana for fruitful discussions.
We are grateful to the referee for valuable comments.
This work was supported by the National Aeronautics and Space Administration under 
grant no.\ NNX13AG92G to D.M. 
We acknowledge the use of computing resources at CIERA funded by NSF PHY-1126812.



\begin{thebibliography}

\bibitem[Aarseth(1999)]{Aarseth1999}
Aarseth, S.  1999, PASP, 111, 1333

\bibitem[Begelman et al.(1980)]{BegelmanBR1980} 
Begelman, M. C., Blandford, R. D., Rees, M. J. 1980, Nature, 287, 307

\bibitem[Berczik et al.(2005)]{BerczikMS2005} 
Berczik, P., Merritt, D., Spurzem, R.  2005, ApJ, 633, 680

\bibitem[Berczik et al.(2006)]{BerczikMSB2006} 
Berczik, P., Merritt, D., Spurzem, R., Bischof, H.  2006, ApJL, 642, L21

\bibitem[Berentzen et al.(2009)]{Berentzen2009} 
Berentzen, I., Preto, M., Berczik, P., Merritt, D., \& Spurzem, R.\  2009, ApJ, 695, 455 

\bibitem[Chatterjee et al.(2003)]{ChatterjeeHL2003} 
Chatterjee, P., Hernquist, L., Loeb, A.  2003, ApJ, 592, 32

\bibitem[Cohn \& Kulsrud(1978)]{CohnKulsrud1978} 
Cohn, H., Kulsrud, R.  1978, ApJ, 226, 1087

\bibitem[Colpi(2014)]{Colpi2014}
Colpi, M.  2014, SSRv, 183, 189

\bibitem[Cui \& Yu(2014)]{CuiYu2014}
Cui, X., Yu, Q.  2014, MNRAS, 437, 777

\bibitem[Dehnen(1993)]{Dehnen1993} 
Dehnen, W.  1993, MNRAS, 265, 250 

\bibitem[Frank \& Rees(1976)]{FrankRees1976}
Frank, J., Rees, M., 1976, MNRAS, 176, 633

\bibitem[Gaburov et al.(2009)]{GaburovHP2009} 
Gaburov, E., Harfst, S., Portegies Zwart, S.  2009, NewA, 14, 630

\bibitem[Gerhard \& Binney(1985)]{GerhardBinney1985}
Gerhard, O., Binney, J.\  1985, MNRAS, 216, 467

\bibitem[Harfst et al.(2008)]{HarfstGMM2008} 
Harfst, S., Gualandris, A., Merritt, D., Mikkola, S.  2008, MNRAS, 389, 2

\bibitem[Hemsendorf et al.(2002)]{HemsendorfSS2002}
Hemsendorf, M., Sigurdsson, S., Spurzem, R.  2002, ApJ, 581, 1256

\bibitem[H\'enon(1971)]{Henon1971}
H\'enon, M., 1971, Ap\&SS, 13, 284

\bibitem[Hernquist \& Barnes(1990)]{HernquistBarnes1990}
Hernquist, L., Barnes, J. 1990, ApJ, 349, 562

\bibitem[Hernquist \& Ostriker(1992)]{HernquistOstriker1992} 
Hernquist, L., Ostriker, J., 1992, ApJ, 386, 375

\bibitem[Holley-Bockelmann \& Sigurdsson(2006)]{HolleySigurdsson2006} 
Holley-Bockelmann, K., Sigurdsson, S.  2006, arXiv:astro-ph/0601520

\bibitem[Holley-Bockelmann \& Khan(2015)]{HolleyKhan2015} 
Holley-Bockelmann, K., Khan, F. M.  2015, arXiv:1505.06203

\bibitem[Holley-Bockelman et al.(2002)]{HolleyBockelmann2002} 
Holley-Bockelmann, K., Mihos, J., Sigurdsson, S., Hernquist, L., Norman, C. 2002, ApJ, 567, 817

\bibitem[Iwasawa et al.(2011)]{Iwasawa2011}
Iwasawa, M., An, S., Matsubayashi, T., Funato, Y., Makino, J., 2011, ApJL, 731, L9

\bibitem[Kandrup et al.(2000)]{KandrupPS2000} 
Kandrup, H., Pogorelov, I., Sideris, I.  2000, MNRAS, 311, 719

\bibitem[Kandrup et al.(2003)]{KandrupSTB2003} 
Kandrup, H., Sideris, I., Terzi{\'c}, B., Bohn, C.  2003, ApJ, 597, 111

\bibitem[Khan et al.(2011)]{KhanJM2011} 
Khan, F. M., Just, A., Merritt, D.  2011, ApJ, 732, 89

\bibitem[Khan et al.(2012)]{Khan2012} 
Khan, F. M., Preto, M., Berczik, P., et al.\ 2012, ApJ, 749, 147 

\bibitem[Khan et al.(2013)]{KhanHBJ2013} 
Khan, F. M., Holley-Bockelmann, K., Berczik, P., Just, A.  2013, ApJ, 773, 100

\bibitem[Li et al.(2014)]{LiHK2014}
Li, B., Holley-Bockelmann, K., Khan, F. M., 2014, ApJ, in press; arXiv:1412.2134

\bibitem[Lightman \& Shapiro(1977)]{LightmanShapiro1977} 
Lightman, A., Shapiro, S.  1977, ApJ, 211, 244

\bibitem[Magorrian \& Tremaine(1999)]{MagorrianTremaine1999} 
Magorrian, J., Tremaine, S.  1999, MNRAS, 309, 447

\bibitem[Makino \& Funato(2004)]{MakinoFunato2004} 
Makino, J., Funato, Y.  2004, ApJ, 602, 93

\bibitem[Meiron \& Laor(2012)]{MeironLaor2012}
Meiron, Y., Laor, A.  2012, MNRAS, 422, 117

\bibitem[Merritt(2001)]{Merritt2001}
Merritt, D.  2001, ApJ, 556, 245

\bibitem[Merritt(2004)]{Merritt2004} 
Merritt, D.  2004, in Coevolution of Black Holes and Galaxies, ed.\ L.~Ho, 
Cambridge Univ.\ press, p.263

\bibitem[Merritt(2006)]{Merritt2006} 
Merritt, D.  2006, ApJ, 648, 976

\bibitem[Merritt(2013)]{MerrittBook} 
Merritt, D., 2013, Dynamics and Evolution of Galactic Nuclei 
(Princeton, NJ: Princeton Univ. Press)

\bibitem[Merritt et al.(2007)]{MerrittMS2007} 
Merritt, D., Mikkola, S., Szell, A.  2007, ApJ, 671, 53

\bibitem[Merritt et al.(2009)]{MerrittSK2009} 
Merritt, D., Schnittman, J., Komossa, S.  2009, ApJ, 699, 1690

\bibitem[Merritt \& Poon(2004)]{MerrittPoon2004} 
Merritt, D., Poon, M.-Y.  2004, ApJ, 606, 788

\bibitem[Merritt \& Quinlan (1998)]{MerrittQuinlan1998}
Merritt, D., Quinlan, G.~D. 1998, ApJ, 498, 625

\bibitem[Merritt \& Szell(2006)]{MerrittSzell2006} 
Merritt, D., Szell, A.,  2006, ApJ, 648, 890

\bibitem[Merritt \& Wang(2005)]{MerrittWang2005} 
Merritt, D., Wang, J. 2005, ApJL, 621, L101

\bibitem[Milosavljevi{\'c} \& Merritt(2001)]{MilosMerritt2001} 
Milosavljevi{\'c}, M., Merritt, D.  2001, ApJ, 563, 34

\bibitem[Milosavljevic \& Merritt(2003a)]{MilosMerritt2003a}
Milosavljevi{\'c}, M., Merritt, D. 2003a, in AIP Conf. Proc., 686, 
The Astrophysics of Gravitational Wave Sources, 
eds. J.~Centrella  and S.~Barnes (Melville, NY: AIP), 201

\bibitem[Milosavljevi{\'c} \& Merritt(2003b)]{MilosMerritt2003b} 
Milosavljevi{\'c}, M., Merritt, D.  2003b, ApJ, 596, 860

\bibitem[Norman \& Silk(1983)]{NormanSilk1983}
Norman, C., Silk, J.\ 1983, ApJ, 266, 502

\bibitem[Peters(1964)]{Peters1964}
Peters, P.C.  1964, PhRvB, 136, 1224

\bibitem[Poon \& Merritt(2002)]{PoonMerritt2002} 
Poon, M.~Y., Merritt, D. 2002, ApJL, 568, L89

\bibitem[Poon \& Merritt(2004)]{PoonMerritt2004} 
Poon, M.~Y., Merritt, D. 2004, ApJ, 606, 774

\bibitem[Preto et al.(2011)]{PretoBBS2011} 
Preto, M., Berentzen, I., Berczik, P., Spurzem, R.  2011, ApJL, 732, L26

\bibitem[Quinlan(1996)]{Quinlan1996} 
Quinlan, G.  1996, NewA, 1, 35

\bibitem[Quinlan \& Hernquist(1997)]{QuinlanHernquist1997} 
Quinlan, G., Hernquist, L.  1997, NewA, 2, 533

\bibitem[Saslaw et al.(1974)]{Saslaw1974}
Saslaw, W.~C., Valtonen, M.~J., Aarseth, S.~J.  1974, ApJ, 190, 253

\bibitem[Sellwood(2015)]{Sellwood2015}
Sellwood, J.  2015, MNRAS, in press; arXiv:1504:06500

\bibitem[Sesana(2010)]{Sesana2010} 
Sesana, A.  2010, ApJ, 719, 851

\bibitem[Sesana \& Khan(2015)]{SesanaKhan2015} 
Sesana, A., Khan, F. M., 2015, arXiv:1505.02062

\bibitem[Sesana et al.(2006)]{SesanaHM2006} 
Sesana, A., Haardt, F., Madau, P.  2006, ApJ, 651, 392

\bibitem[Sesana et al.(2007)]{SesanaHM2007} 
Sesana, A., Haardt, F., Madau, P.  2007, ApJ, 660, 546

\bibitem[Sesana et al.(2008)]{SesanaHM2008} 
Sesana, A., Haardt, F., Madau, P.  2008, ApJ, 686, 432

\bibitem[Sesana et al.(2011)]{SesanaGD2011} 
Sesana, A., Gualandris, A., Dotti, M.  2011, MNRAS, 415, L35

\bibitem[Spitzer \& Hart(1971)]{SpitzerHart1971}
Spitzer, L., Hart, M.  1971, ApJ, 164, 399

\bibitem[Vasiliev(2014a)]{Vasiliev2014a}
Vasiliev, E.  2014a, CQGra, 31, 244002

\bibitem[Vasiliev(2014b)]{Vasiliev2014b} 
Vasiliev, E.  2014b, arXiv:1411.1762

\bibitem[Vasiliev(2015)]{Vasiliev2015} 
Vasiliev, E.  2015, MNRAS, 446, 3150

\bibitem[Vasiliev \& Merritt(2013)]{VasilievMerritt2013} 
Vasiliev, E., Merritt, D.  2013, ApJ, 774, 87

\bibitem[Vasiliev et al.(2014)]{VasilievAM2014} 
Vasiliev, E., Antonini, F., Merritt, D.  2014, ApJ, 785, 163

\bibitem[Weinberg(1996)]{Weinberg1996}
Weinberg, M.  1996, ApJ, 470, 715

\bibitem[Weinberg(1998)]{Weinberg1998}
Weinberg, M.  1998, MNRAS, 297, 101

\bibitem[Yu(2002)]{Yu2002} 
Yu, Q.  2002, MNRAS, 331, 935

\end{thebibliography}
\end{document}